\documentclass[runningheads,a4paper]{llncs}

\usepackage[T1]{fontenc}
\usepackage{amsmath}
\usepackage{amsthm}
\usepackage{amssymb}
\usepackage{mathtools}
\usepackage{url}
\usepackage{hyperref}
\usepackage[utf8]{inputenc}
\usepackage{bussproofs}
\usepackage{xparse}
\usepackage{stmaryrd}
\usepackage{xspace}
\usepackage{framed}
\usepackage{booktabs}
\usepackage{csquotes}
\usepackage{xcolor}
\usepackage{listings}
\usepackage{tikz}
\usepackage{algorithm}
\usepackage{algpseudocode}
\usetikzlibrary{positioning}
\usepackage{tkz-graph}
\usetikzlibrary{calc}
\usetikzlibrary{arrows}
\usepackage{relsize}
\usepackage{enumerate}
\usepackage{multirow}
\usepackage{subcaption} 
\usepackage{appendix}

\usepackage{thmtools}
\usepackage{thm-restate}
\usepackage{hyperref}
\usepackage{cleveref}

\theoremstyle{definition}
\newtheorem{requirement}{Requirement}

\usepackage{enumitem}
\setlist[itemize]{noitemsep, topsep=0.2pt}

\usepackage[outercaption]{sidecap}    

\makeatletter
\newcommand{\nobrackettag}[0]{\def\tagform@##1{\maketag@@@{##1}}}
\makeatother

\newcommand{\A}{\mathcal{A}}
\newcommand{\B}{\mathcal{B}}

\newcommand{\I}{\mathcal{I}}

\newcommand{\G}{\mathcal{G}}

\newcommand{\K}{\mathcal{K}}
\renewcommand{\L}{\mathcal{L}}

\newcommand{\R}{\mathcal{R}}
\renewcommand{\S}{\mathcal{S}}
\newcommand{\T}{\mathcal{T}}

\newcommand{\X}{\mathcal{X}}

\newcommand{\x}{\mathbf{x}}

\newcommand{\tup}[1]{\langle #1\rangle}            

\newcommand{\dlliter}{\textit{DL-Lite}\ensuremath{_{\R}}\xspace}

\newcommand{\lcan}{\L_{\can}}

\newcommand{\PSPACE}{\textsc{PSpace}\xspace}

\newcommand{\NP}{\textsc{NP}\xspace}
\newcommand{\coNP}{\textsc{coNP}\xspace}

\newcommand{\sigmaptwo}{$\mathsf{\Sigma}^{\mathsf{P}}_2$\xspace}

\newcommand{\stt}[1]{\small{\texttt{#1}}}

\newcommand{\ignore}[1]{}

\newcommand{\restRangeIn}[2]{#1 \triangleright #2}
\newcommand{\restRange}[2]{#1 \blacktriangleright #2}

\newcommand{\JOIN}{\textsc{join}\xspace}
\newcommand{\UNION}{\textsc{union}\xspace}

\newcommand{\OPT}{\textsc{opt}\xspace}

\newcommand{\SELECT}{\textsc{select}\xspace}

\newcommand{\sparql}{\textsc{sparql}\xspace}
\newcommand{\rdf}{\textsc{rdf}\xspace}
\newcommand{\rdfs}{\textsc{rdfs}\xspace}
\newcommand{\owl}{\textsc{owl}\xspace}

\newcommand{\owltwoql}{\textsc{owl\,2\,ql}\xspace}

\newcommand{\ei}{\emph{(i)}\xspace}
\newcommand{\eii}{\emph{(ii)}\xspace}

\newcommand{\modelsOf}{\mathsf{mod}}
\newcommand{\dom}{\mathsf{dom}}
\newcommand{\range}{\mathsf{range}}
\newcommand{\vars}{\mathsf{vars}}
\newcommand{\ans}{\mathsf{ans}}
\newcommand{\certAns}{\mathsf{certAns}}
\newcommand{\cCertAns}{\mathsf{eRAns}}
\newcommand{\eCertAns}{\mathsf{eCertAns}}
\newcommand{\sparqlAns}{\mathsf{sparqlAns}}
\newcommand{\eAns}{\mathsf{eAns}}
\newcommand{\canAns}{\mathsf{canAns}}
\newcommand{\certCanAns}{\mathsf{mCanAns}}
\newcommand{\aDom}{\mathsf{aDom}}
\newcommand{\can}{\mathsf{can}}
\newcommand{\adm}{\mathsf{adm}}
\newcommand{\branch}{\mathsf{branch}}
\newcommand{\restCanAns}{\mathsf{restCanAns}}
\newcommand{\base}{\mathsf{base}}

\newcommand{\NI}{\mathsf{N}_\mathsf{I}}
\newcommand{\NR}{\mathsf{N}_\mathsf{R}}

\newcommand{\NC}{\mathsf{N}_\mathsf{C}}
\newcommand{\NV}{\mathsf{N}_\mathsf{V}}

\newcommand{\U}{\mathsf{U}}

\title{Certain Answers to a \sparql Query over a Knowledge Base (extended version)}
\author{
Julien Corman\inst{1}
\and Guohui Xiao\inst{1}
}

\authorrunning{Julien Corman, Guohui Xiao}
\titlerunning{Certain Answers to a \sparql Query over a Knowledge Base}
\institute{
Free University of Bozen-Bolzano, Bolzano, Italy 
}

\begin{document}
\maketitle



\begin{abstract}
Ontology-Mediated Query Answering (OMQA) is a well-established framework to answer queries over an \rdfs or \owl Knowledge Base (KB).
OMQA was originally designed for unions of conjunctive queries (UCQs),
and based on \emph{certain answers}.
More recently, OMQA has been extended to \sparql queries, but to our knowledge, none of the efforts made in this direction (either in the literature,
or the so-called \sparql \emph{entailment regimes}) is able to capture both certain answers for UCQs and the standard interpretation of \sparql over a plain graph.
We formalize these as requirements to be met by any semantics aiming at conciliating certain answers and \sparql answers,
and define three additional requirements, which generalize to KBs some basic properties of \sparql answers.
Then we show that a semantics can be defined that satisfies all requirements for \sparql queries with \stt{SELECT}, \stt{UNION}, and \stt{OPTIONAL},
and for DLs with the canonical model property.
We also investigate combined complexity for query answering under such a semantics over \dlliter KBs.
In particular,
we show for different fragments of \sparql that known upper-bounds for query answering over a plain graph are matched.
\end{abstract}


\sloppy


\section{Introduction}
\label{sec:introduction}
\sparql is an expressive SQL-like query language designed for Semantic Web data, exposed as \rdf graphs.
Recently,
\sparql has been extended with so-called \emph{entailment regimes},
which specify different semantics to query an \rdfs or \owl \emph{Knowledge Base} (KB),
i.e. data enriched with a background theory.
This allows retrieving answers to a query not only over the facts explicitly stated in the KB,
but more generally over what can be inferred from the KB. 

The \sparql entailment regimes are in turn largely influenced by theoretical work on \emph{Ontology Mediated Query Answering} (OMQA),
notably in the field of \emph{Description Logics} (DLs).
However, OMQA was initially developed for \emph{unions of conjunctive queries} (UCQs),
which have a limited expressivity when compared to \sparql.
 It turns out that conciliating the standard (compositional) semantics of \sparql on the one hand,
and the semantics used for OMQA on the other hand,
called \emph{certain answers},
is non-trivial.

As an illustration,
Example~\ref{ex:cq} provides a simple KB and \sparql query.
The dataset (a.k.a \emph{ABox}) $\A$ states that \stt{Alice} is a driver,
whereas the background theory (a.k.a. \emph{TBox}) $\T$ states that a driver must have a license
(for conciseness, we use DLs for the TBox, rather than some concrete syntax of \owl).
Finally, the \sparql query $q$ retrieves all individuals that have a license.




\begin{example}\label{ex:cq}\ \\
  $\A = \{\stt{Driver(Alice)}\}$\\
  $\T = \{\stt{Driver} \sqsubseteq \exists\stt{hasLicense}\}$\\
  $q\ = \stt{SELECT ?x WHERE \{ ?x hasLicense ?y \}}$
\end{example}

Intuitively, one expects \stt{Alice} to be retrieved as an answer to $q$.
And it would indeed be the case under certain answer semantics,
if one considers the natural translation of this query into a UCQ.
On the other hand,
under the standard semantics of \sparql 1.1~\cite{sparql11},
this query has no answer.
This is expected,
since the fact that \stt{Alice} has a driving license is not present in the ABox.
More surprisingly though,
under all \sparql entailment regimes~\cite{sparql-regimes},
this query also has no answer.

This mismatch between certain answers and entailment regimes has already been discussed in depth 
in~\cite{ahmetaj_towards_2015},
where the interpretation of the \stt{OPTIONAL} operator of \sparql is identified as a challenge,
when trying to define a suitable semantics for \sparql that complies with certain answers for UCQs.
A concrete proposal is also made in~\cite{ahmetaj_towards_2015} in this direction.
 Unfortunately,
this semantics does not comply with the standard semantics of \sparql when the TBox is empty.
This means that a same query over a plain \rdf graph may yield different answers,
depending on whether it is evaluated under this semantics,
or under the one defined in the \sparql~1.1 specification~\cite{sparql11}.

 We propose in this article to investigate whether and how this dilemma can be solved,
 for the so-called \emph{set semantics} of \sparql and certain answers.
To this end, we first formulate in Section~\ref{sec:requirements} some \emph{requirements} to be met by any reasonable semantics meant to conciliate certain answers and standard \sparql answers.
Then in Section~\ref{sec:semantics},
we use these requirements to review different semantics.
We also show that all requirements can be satisfied,
for the fragment of \sparql with \stt{SELECT}, \stt{UNION} and \stt{OPTIONAL},
and for KBs that admit a unique \emph{canonical model}.
Finally, in Section~\ref{sec:complexity},
we provide combined complexity results for query answering under this semantics,
over KBs in \dlliter,
one of the most popular DLs tailored for query answering,
which correspond to the \owltwoql standard.
We show in particular that upper bounds for this problem match results already known to hold for \sparql over plain graphs,
which means that under this semantics,
and as far as worst-case complexity is concerned,
the presence of a TBox does not introduce a computational overhead. 
Before this,
Section~\ref{sec:prelim} introduces preliminary notions,
and Section~\ref{sec:sparql-over-kb} reviews existing semantics for \sparql over a KB.
Proofs can be found in apppendix.






\section{Preliminaries}
\label{sec:prelim}

We assume countably infinite and mutually disjoint sets $\NI$, $\NC$,
$\NR$, and $\NV$ of \emph{individuals} (constants), \emph{concept names} (unary predicates),
\emph{role names} (binary predicates), and variables respectively.
We also assume a countably infinite universe $\U$, such that $\NI \subseteq \U$.
For clarity, we abstract away from concrete domains (as well as \rdf term types),
since these are irrelevant to the content of this paper.
We also assume that $\NI$, $\NC$ and $\NR$ do not contain any reserved term from the \rdf/\rdfs/\owl vocabularies (such as \stt{rdfs:subClassOf}, \stt{owl:disjointWith}, etc.)

\subsection{\rdf and \sparql}\label{sec:sparql}

An (\rdf) \emph{triple} is an element of
$(\NI \times \{\stt{rdf:type}\} \times \NC) \cup (\NI \times \NR
\times \NI)$.  An \rdf graph $\A$ is a set of triples.
For the concrete syntax of \sparql, we refer to the
specification~\cite{sparql11}.
Following~\cite{ahmetaj_towards_2015},
we focus on \sparql queries whose triple patterns are either in
$(\NV \cup \NI) \times \{\stt{rdf:type}\} \times \NC$, or in
$(\NV \cup \NI) \times \NR \times (\NV \cup \NI)$.
For readability,
we represent triples and triple patterns as atoms in prefix notation, i.e. we
use $A(t)$ rather than $(t,\stt{rdf:type}, A)$ and for $r \in \NR$, we
use $r(t_1,t_2)$ rather than $(t_1,r,t_2)$.  If $q$ is a \sparql
query, we use $\vars(q)$ to denote the set of variables projected by
$q$.

We adopt (roughly) the abstract syntax provided in~\cite{perez_semantics_2009} for the fragment of \sparql with the \stt{SELECT},
\stt{UNION} and \stt{OPTIONAL} operators,
using the following grammar,
where $t$ is a \sparql triple pattern,
and $X \subseteq \NV$:
$$q \quad ::= \quad t \mid \SELECT_X\  q \mid q\ \UNION\ q \mid q\ \JOIN\ q \mid  q\ \OPT\ q$$
In addition,
if $q = \SELECT_X\  q'$,
then $X \subseteq \vars(q')$ must hold.
In order to refer to fragments of this language,
we use the letters S, U, J and O (in this order),
for \SELECT, \UNION, \JOIN, and \OPT respectively.
E.g. ``SUJO'' stands for the full language,
``UJ'' for the fragment with \UNION and \JOIN only, etc.

If $\omega$ is a function,
we use $\dom(\omega)$ (resp. $\range(\omega)$) to designate its domain (resp. range).
Two functions $\omega_1$ and $\omega_2$ are \emph{compatible},
denoted with $\omega_1 \sim \omega_2$,
iff $\omega_1(x) = \omega_2(x)$ for each $x \in \dom(\omega_1) \cap \dom(\omega_2)$.
If $\omega_1$ and $\omega_2$ are compatible,
then $\omega_1 \cup \omega_2$ is the only function with domain $\dom(\omega_1) \cup \dom(\omega_2)$ that is compatible with $\omega_1$ and $\omega_2$.
We say that a function $\omega_2$ \emph{extends} a function $\omega_1$,
noted $\omega_1 \preceq \omega_2$,
iff $\dom(\omega_1) \subseteq \dom(\omega_2)$ and $\omega_1 ~\sim \omega_2$.
Finally, we use $\omega|_X$ (resp. $\omega\|_X$) to designate the restriction of function $\omega$ to domain (resp. co-domain) $X$,
i.e. $\omega|_X$ is the only function compatible with $\omega$ that verifies $\dom(\omega|_X) = \dom(\omega) \cap X$,
and $\omega\|_X$ is the only function compatible with $\omega$ that verifies $\dom(\omega\|_X) = \{v \in \dom(\omega) \mid \omega(v) \in X\}$.

A \emph{solution mapping} is a function from a finite subset of $\NV$ to $\U$.
If $\Omega_1$ and $\Omega_2$ are sets of solutions mappings and $X \subseteq V$, then:
\begin{center}
\begin{tabular}{lll}
$\Omega_1 \bowtie \Omega_2$& $=$& $\{ \omega_1 \cup \omega_2 \mid (\omega_1, \omega_2) \in \Omega_1 \times \Omega_2 \textnormal{ and } \omega_1 \sim \omega_2 \}$\\
$\Omega_1 \setminus \Omega_2$& $=$ &$\{ \omega_1 \mid \omega_1 \in \Omega_1 \textnormal{ and } \omega_1 \not\sim \omega_2 \textnormal{ for all } \omega_2 \in \Omega_2\}$\\
$\pi_X \Omega$& $=$& $\{ \omega|_X \mid \omega \in \Omega\}$
\end{tabular}
\end{center}


If $q$ is a \sparql query and $\omega$ a solution mapping s.t. $\vars(q) \subseteq \dom(\omega)$,
we use $\omega(q)$ to designate the query identical to $q$,
but where each occurrence of variable $x$ in a triple pattern is replaced by $\omega(x)$.


We now reproduce the inductive definition of answers to a \sparql query $q$ over a graph $\A$,
denoted $\sparqlAns(q, \A)$,
provided in~\cite{perez_semantics_2009} for the SUJO fragment (and for set semantics).
 \begin{definition}[\sparql answers over a plain graph~\cite{perez_semantics_2009}]\label{def:sparqlAns}
\begin{center}
  \begin{tabular}{lll}
    \multicolumn{3}{l}{If $q$ is a triple pattern, then $\sparqlAns(q,\A) = \{\omega \mid \dom(\omega) = \vars(q) \textnormal{ and } \omega(q) \in \A\}$}\\
   $\sparqlAns(q_1\ \UNION\ q_2, \A)$& $=$& $\sparqlAns(q_1, \A) \cup \sparqlAns(q_2, \A)$\\
   $\sparqlAns(q_1\ \JOIN\ q_2, \A)$& $=$& $ \sparqlAns(q_1, \A) \bowtie \sparqlAns(q_2, \A)$\\
    $\sparqlAns(q_1\ \OPT\ q_2, \A) $&$=$&$(\sparqlAns(q_1,\A) \bowtie \sparqlAns(q_2,\A))\ \cup $\\
                                           &&$(\sparqlAns(q_1,\A) \setminus  \sparqlAns(q_2,\A))$\\
   $\sparqlAns(\SELECT_X\ q, \A) $&$=$&$ \pi_X \sparqlAns(q, \A)$\\
\end{tabular}
    \end{center}
 \end{definition}


\subsection{Description Logic KB,  UCQs and Certain Answers}
\label{sec:KB}

As is conventional in the Description Logics (DL) literature, we
represent a KB $\K$ as a pair $\K = \tup{\T,\A}$,
where $\A$ is called the \emph{ABox} of $\K$,
which contains assertions about individuals,
and $\T$ is called the \emph{TBox} of $\K$,
which contains more abstract knowledge.
An ABox is a finite set of atoms of the form $A(c)$ or $r(c_1, c_2)$,
where $A \in \NC$, $r \in \NR$ and $c,c_1,c_2 \in \NI$.
A TBox is a finite set of logical \emph{axioms}, whose form depends on the particular DL.
For a KB $\K = \tup{\T,\A}$,
 the \emph{active domain} of $\K$,
denoted with $\aDom(\K)$,
is the set of elements of $\NI$ that appear (syntactically) in $\T$ or~$\A$.


The semantics of DL KBs is defined in terms of (first-order) \emph{interpretations}.
We adopt in this article the \emph{standard name assumption}:
an interpretation is a structure $\I = \tup{\Delta^\I, \cdot^\I}$,
where the \emph{domain} $\Delta^\I$ of $\I$ is a non-empty subset of $\U$,
and the \emph{interpretation function} $\cdot^I$ of $\I$ maps each $c \in \NI$ to itself,
and each $A \in \NC$ (resp. $r \in \NR$) to a unary (resp, binary) relation $A^I$ (resp. $r^\I$) over $\Delta^\I$.
An interpretation $\I$ is a \emph{model} of a KB $\K = \tup{\T,\A}$
if it satisfies every assertion in $\A$ and axiom in $\T$.  For the
formal definition of ``satisfies'', we refer to~\cite{dl-handbook}.
%

%

If $\K$ is a KB, we use $\modelsOf(\K)$ to denote the set of models of $\K$.
We focus on \emph{satisfiable} KBs only,
i.e. KBs that admit at least one model,
since any formula can be trivially derived from an unsatisfiable KB.
We also omit this precision for readability.
So ``any KB'' below is a shortcut for ``any satisfiable KB''.

For a DL KB $\K$, an interpretation $\I_c \in \modelsOf(\K)$ is a \emph{canonical model} of $\K$ if $\I_c$ can be homomorphically mapped to any  $\I \in \modelsOf(\K)$.
We say that a DL $\L$ has the \emph{canonical model property} if every KB in $\L$ has a \emph{unique} canonical model up to isomorphism.
This is a key property of DLs tailored for query answering,
and many DLs,
e.g. $\dlliter$, $\mathcal{EL}$ or Horn-$\mathcal{SHIQ}$, have this property.

An interpretation (or an ABox) can also be viewed as a (possibly infinite) \rdf graph,
with triples $ \{A(d) \mid d \in A^\I, A \in \NC \} \cup \{ r(d_1, d_2) \mid
(d_1,d_2)\in r^\I, r \in \NR \}$.
This is a slight abuse (the \rdf standard does not admit infinite graphs), 
but we will nonetheless use this convention throughout the article,
in order to simplify notation.

%




A \emph{conjunctive query} (CQ) $h$ is a expression of the form:
  $$h(\x)\gets p_1(\x_1), \ldots, p_m(\x_m)$$
  where $h, p_i$ are predicates and $\x, \x_i$ are tuple over $\NV$.
Abusing notation, we may use $\x$ (resp. $\x_i$) below to designate the elements of $\x$ (resp. $\x_i$) viewed as a set.
An additional syntactic requirement on a CQ is that $\x \subseteq \x_1 \cup .. \cup \x_m$.
The variables in $\x$ are called \emph{distinguished},
and we use $\vars(h)$ to designate the distinguished variables of CQ $h$.
  We focus in this article on CQs where each $p_i$ is unary or binary,
  i.e. $p_i \in \NC \cup \NR$.
A \emph{match} for $h$ in an interpretation $\I$ is a total function $\rho$ from $\x_1 \cup \ldots \cup \x_m$ to $\Delta^{\I}$ such that $\rho(\x_i)\,{\in}\,(p_i)^{\I}$
for $i \in \{1 .. m\}$.
A mapping $\omega$ is an \emph{answer} to $h$ over $\I$ iff there is a match $\rho$ for $h$ in $\I$ s.t. $\omega = \rho|_{\vars(h)}$.

A union of conjunctive queries (UCQ) is a set $q = \{h_1, \ldots, h_n\}$ of CQs sharing the same distinguished variables,
and $\omega$ is an \emph{answer} to $q$ over $\I$ iff $\omega$ is an answer to some $h_i$ over $\I$.
Finally,
$\omega$ is a \emph{certain answer} to $q$ over a KB $\K$ iff $\range(\omega) \subseteq \aDom(\K)$ and $\omega$ is an answer to $q$ over each $\I \in \modelsOf(\K)$.
We use $\certAns(q,\K)$ to designate the set of certain answers to $q$ over $\K$.

CQs and UCQs have a straightforward representation as \sparql queries.
The CQ $h(\x)\gets p_1(\x_1), \ldots, p_m(\x_m)$ in \sparql syntax is written:
$$\SELECT_\x\  (p_1(\x_1) \ \JOIN\ ..\ \JOIN\ p_m(\x_m))$$
And a UCQ in \sparql syntax is of the form:
$$h_1\ \UNION\ ..\ \UNION\ h_n$$
where each $h_i$ is a CQ in \sparql syntax,
and $\vars(h_i) = \vars(h_j)$ for $i,j \in \{1..n\}$.

\clearpage
\section{Querying a DL KB with \sparql: Existing Semantics}
\label{sec:sparql-over-kb}

In this section, we review existing semantics for \sparql over a DL KB.
We start by briefly recalling some features of the W3C specification for the \sparql 1.1 entailment regimes~\cite{sparql-regimes}.
This specification defines different ways to take into account the semantics of \rdf, \rdfs or \owl,
in order to infer additional answers to a \sparql query.
We ignore the aspects pertaining to querying blank nodes and concept/role names,
which fall out of the scope of this paper,
and focus on the entailment regimes parameterized by an \owl profile,
i.e. a DL $\L$.
In short, the $\L$-entailment regime modifies the evaluation of a \sparql query $q$ over an $\L$-KB
$\K = \tup{\T,\A}$ as follows:
\begin{enumerate}
\item Triple patterns are not evaluated over the ABox $\A$,
  but instead over the so-called \emph{entailed graph},
  which consists of all ABox assertions entailed by $\K$.
  This includes assertions of the form $C(a)$,
  where $C$ is a complex concept expression allowed in $\L$.
  The semantics of other \sparql operators is preserved.
\item The \sparql query can use $\L$-concepts in triple pattern, e.g. $\exists\stt{hasLicense(x)}$.
\end{enumerate}
Consider again Example~\ref{ex:cq}.
under the \owl2 QL entailment regime for instance,
which corresponds (roughly) to the DL \dlliter.
In this example,
the query $\exists\stt{hasLicense}(x)$ has $\{x \mapsto\stt{Alice}\}$ as unique answer:
since the entailed graph contains all ABox assertions entailed by $\K$,
it contains the assertion $\exists\stt{hasLicense}(\stt{Alice})$ (again, we use the DL syntax rather than \owl,
for readability).
  
  So the expressivity of the $\L$-entailment regime is limited by the concepts that can be expressed in $\L$.
  This is why~\cite{kontchakov_answering_2014} proposed to extend the semantics of the \owl2 QL profile,
  retrieving instances of concepts that cannot be expressed in \dlliter (e.g. concepts of the form $\exists r_1.\exists r_2$).
  Still,
  under this semantics as well as all entailment regimes defined in the specification,
  the query $\SELECT_{\{x\}} \stt{hasLicense}(x, y)$ has no answer over the KB of Example~\ref{ex:cq},
because the entailed graph does not contain any assertion of the form $\stt{hasLicense}(\stt{Alice}, e)$.

    This point was discussed in depth in~\cite{ahmetaj_towards_2015},
    for the SUJO fragment,
    and based on remarks made earlier in~\cite{arenas_querying_2011}.
    The current paper essentially builds upon this discussion,
    which is why we reproduce it below. 
A first remark made in~\cite{arenas_querying_2011} and \cite{ahmetaj_towards_2015} is that the \OPT operator of \sparql prevents the usage of certain answers,
even when querying a plain graph (or equivalently, a KB with empty TBox).
This can be seen with Example~\ref{ex:opt_ca}.

\begin{restatable}{example}{exampleOptCA}\label{ex:opt_ca}\ \\
  $\A = \{\stt{Person(Alice)}\}$\\
  $q\ = \stt{Person}(x)\  \OPT\ \stt{hasLicense}(x,y)$
\end{restatable}

In this example, according to the \sparql specification,
the mapping $\omega = \{x \mapsto \stt{Alice}\}$ is the only answer to $q$ over $\A$,
i.e. $\sparqlAns(q,\A) = \{\omega\}$.
But $\omega$ is not a certain answer to $q$ over the KB $\tup{\emptyset,\A}$.
Consider for instance the interpretation $\I$ defined by $\I = \A \cup \{\stt{hasLicense}(\stt{Alice}, 12345)\}$.
Then $\sparqlAns(q,\I) = {\{\{x \mapsto \stt{Alice}, y \mapsto 12345\}\}}$.
So $\omega \not\in \certAns(q,\tup{\emptyset,\A})$.

Then in~\cite{arenas_querying_2011} and~\cite{ahmetaj_towards_2015} still,
the authors remark that in this example,
$\omega$ can nonetheless be \emph{extended} to an answer in every model of $\tup{\emptyset,\A}$.
This is the main intuition used in~\cite{ahmetaj_towards_2015} to adapt the definition of certain answers to \sparql queries with \OPT.
If $q$ is a query and $\I$ an interpretation,
let $\eAns(q,\I)$ designate all mappings that can be extended to an answer to $q$ in $\I$,
i.e.:
 $$\eAns(q,\I) = \{\omega \mid \omega \preceq \omega' \textnormal{ for some } \omega' \in \sparqlAns(q,\I)\}$$ 
Then if $\K$ is a KB,
the set $\eCertAns(q,\K)$ of mappings that can be extended to an answer in every model of $\K$ is defined as:
 $$\eCertAns(q,\K) = \bigcap\limits_{\I \in \modelsOf(\K)} \eAns(q,\I)$$
But as pointed out in~\cite{ahmetaj_towards_2015},
$\eCertAns(q,\I)$ does not comply with \sparql answers over a plain graph (i.e. when the TBox is empty).
Indeed,
if some $\omega$ can be extended to an answer in every model of the KB,
then this is also the case of any mapping that $\omega$ extends (e.g. trivially the empty mapping).
So in Example~\ref{ex:opt_ca},
$\eCertAns(q,\tup{\emptyset,\A}) = \{\{\}, {\{x \mapsto \stt{Alice}\}}\}$,
whereas $\sparqlAns(q,\A) = \{\{x \mapsto \stt{Alice}\}\}$.

The semantics proposed in~\cite{ahmetaj_towards_2015} is designed to solve this issue.
The precise scope of the proposal is so-called \emph{well-designed} SUJO queries (see~\cite{perez_semantics_2009} for a definition),
in some normal form (no \UNION in the scope of \SELECT, \JOIN or \OPT,
no \SELECT in the scope of \JOIN or \OPT, and no \OPT in the scope of \JOIN).\footnote{
  This is without loss of expressivity, but normalization may cause an exponential blowup.}
Given a KB $\K$,
the solution consists in retaining,
for each maximal SJO subquery $q'$,
the \emph{maximal} elements of $\eCertAns(q',\K)$ w.r.t $\preceq$.
An additional restriction is put on the domain of such solution mappings,
based on the so-called \emph{pattern-tree} representation (defined in~\cite{letelier_static_2013}) of well-designed SJO queries.
The \UNION operator on the other hand is evaluated compositionally, as in Definition~\ref{def:sparqlAns}.

But as illustrated by the authors,
this proposal does not comply with the standard semantics for \sparql over plain graphs.
Example~\ref{ex:ahmetaj} below reproduces the one given in~\cite[Example 4]{ahmetaj_towards_2015}:
\begin{example}\label{ex:ahmetaj}\ \\
$\A = \{
  \stt{teachesTo}(\stt{Alice},\stt{Bob}),
  \stt{knows}(\stt{Bob}, \stt{Carol}),
  \stt{teachesTo}(\stt{Alice}, \stt{Dan})
  \}$ \\
  $q\ = \SELECT_{\{x,z\}} (\stt{teachesTo}(x, y)\ \OPT\ \stt{knows}(y,z))$
\end{example}
In this example,
$\sparqlAns(q,\A) = \{\{x \mapsto \stt{Alice}, z \mapsto \stt{Carol}\}, \{x \mapsto \stt{Alice}\}\}$.
Instead, the semantics proposed in~\cite{ahmetaj_towards_2015} yields $\{\{x \mapsto \stt{Alice}, z \mapsto \stt{Carol}\}\}$.

Section~\ref{sec:certCanAns} below defines a different semantics for evaluating a \sparql query over a KB,
which coincides not only with certain answers for UCQs (as opposed to the \sparql entailment regimes and~\cite{kontchakov_answering_2014}),
but also with the \sparql specification in the case where the TBox is empty (as opposed to the proposal made in \cite{ahmetaj_towards_2015}).

Before continuing, other works need to be mentioned,
even though they are not immediately related to the problem addressed in this paper.
First, a modification of the entailment regimes' semantics was proposed in~\cite{kostylev_semantics_2014} for the SJO fragment extended with the \sparql \stt{FILTER} operator.
For DLs with negation,
it consists in ruling out a partial solution mappings if it cannot be extended to an answer in any model of the KB.
Finally, another topic of interest when it comes to \sparql and certain answers,
but which falls out of the scope of this paper,
is the treatment of \emph{blank nodes},
discussed in the specification of \sparql entailment regimes~\cite{sparql-regimes},
and more recently in~\cite{gutierrez2016certain} and~\cite{hernandez2018certain}.


\section{Requirements}
\label{sec:requirements}

As seen in the previous section,
existing semantics for \sparql answers over a KB fail to comply either with certain answers (for the fragment of \sparql that corresponds to UCQs),
or with \sparql answers over a plain graph when the TBox is empty.

We will show in Section~\ref{sec:semantics} that these two requirements are compatible for some DLs and fragments of \sparql. 
But first, in this section,
we formalize these two requirements,
as properties to met by any semantics whose purpose is to conciliate certain answers and \sparql answers. 
We also define three additional requirements (called \emph{\OPT extension}, \emph{variable binding} and \emph{binding provenance}),
which generalizes to KBs some basic properties of \sparql answers over plain graphs.
We note that these requirements apply to arbitrary DLs,
whereas Section~\ref{sec:semantics} focuses instead on specific families of DLs.

If $q$ is a \sparql query and $\K$ a KB,
we use $\ans(q,\K)$ below to denote the answers to $q$ over $\K$ under some (underspecified) semantics.
This allows us to define properties to be met by such a semantics.

Requirement~\ref{req:ca_comp} states that $\ans(q,\K)$ should coincide with certain answers for UCQs.
\begin{requirement}[Certain answer compliance]\label{req:ca_comp} For any UCQ $q$ and KB $\K$,
  $$\ans(q,\K) = \certAns(q,\K)$$
\end{requirement}

Requirement~\ref{req:sa_comp} corresponds to the limitation of~\cite{ahmetaj_towards_2015} identified in Section~\ref{sec:sparql-over-kb}.
It requires that $\ans(q,\tup{\emptyset,\A})$ coincide with answers over $\A$, as defined in the \sparql specification.
\begin{requirement}[\sparql answer compliance]\label{req:sa_comp}
  For any query $q$ and ABox $\A$,
  $$\ans(q,\tup{\emptyset,\A}) = \sparqlAns(q,\A)$$
\end{requirement}

As will be seen in the next section,
it is possible to define semantics that verify Requirements~\ref{req:ca_comp} and~\ref{req:sa_comp},
but fail to comply with basic properties of \sparql answers over a plain graph.
This is why we define additional requirements.

First, as observed in \cite{kostylev_semantics_2014} for instance,
the \OPT operator of \sparql was introduced to ``not reject the solutions because some part of the query pattern does not match''~\cite{sparql11}.
Or in other words, for each answer $\omega$ to the left operand of an \OPT,
either $\omega$ or some extension of $\omega$ is expected be present in the answers to the whole expression.  
Let $\preceq_g$ be the partial order over sets of solution mappings defined by $\Omega_1 \preceq_g \Omega_2$ iff,
for each $\omega_1 \in \Omega_1$,
there is a $\omega_2 \in \Omega_2$ s.t. $\omega_1 \preceq \omega_2$.
Then this property is expressed with Requirement~\ref{req:op_mon}.
\begin{requirement}[\OPT extension]\label{req:op_mon}
  For any queries $q_1, q_2$ and KB $\K$:
  $$\ans(q_1, \K) \preceq_g \ans(q_1\ \OPT\ q_2,\K)$$
\end{requirement}

Another important property of \sparql answers over plain graphs pertains to bound variables.
Indeed,
a \sparql query $q$ (with \UNION and/or \OPT) may allow \emph{partial} solution mappings,
i.e. whose domain does not cover all variables projected by $q$.
For instance,
in Example~\ref{ex:opt_ca},
$\omega = \{x \mapsto \stt{Alice}\} \in \sparqlAns(q,\A)$,
even though the variables projected by $q$ are $x$ and $y$.
In such a case,
we say that variable $x$ is \emph{bound} by $\omega$, whereas variable $y$ is not.
Then a \sparql query may only admit answers that bind certain sets of variables.
For instance the query
$A(x)\ \OPT\ (R(x,y)\ \JOIN\ R(y,z))$ admits answers that bind either $\{x\}$ or $\{x, y, z\}$.
But it does not admit answers that bind another set of variables ($\{y\}$,$\{x,y\}$, etc.).
So a natural requirement when generalizing \sparql answers to KBs is to respect such constraints.
We say that a set $X$ of variables is \emph{admissible} for a query $q$
iff there exists a graph $\A$ and solution mapping $\omega$ s.t.
$\omega \in \sparqlAns(q,\A) \textnormal{ and } \dom(\omega) = X$.
Unfortunately, for queries with \stt{OPTIONAL},
whether a given set of variables is admissible for a given query is undecidable.
So we adopt instead a relaxed notion of admissible bindings.
For a SUJO query $q$,
we use $\adm(q)$ to denote the family of sets of variables defined inductively as follows:
\newpage
\begin{definition}[Definition of $\adm(q)$ for the SUJO fragment]~\label{def:adm_ind}
\begin{center}
  \begin{tabular}{lll}
    \multicolumn{3}{l}{If $q$ is a triple pattern, then $\adm(q) = \{\vars(q)\}$}\\
    $\adm(\SELECT_X\ q)$ &$= \{\ X' \cap X$&$\mid X' \in \adm(q)\ \}$\\
    $\adm(q_1\ \JOIN\ q_2)$ &$= \{\ X_1 \cup X_2\ $&$\mid (X_1,X_2) \in \adm(q_1) \times \adm(q_2)\ \}$\\
    $\adm(q_1\ \OPT\ q_2)$ &\multicolumn{2}{l}{$= \adm(q_1) \cup \adm(q_1\ \JOIN\ q_2)$}\\
    $\adm(q_1\ \UNION\ q_2)$ &\multicolumn{2}{l}{$= \adm(q_1) \cup \adm(q_2)$}
\end{tabular}
\end{center}
\end{definition}
We can now formulate the corresponding requirement:
\begin{requirement}[Variable binding]\label{req:var_bind}
  For any SUJO query $q$, KB $\K$ and $\omega \in \ans(q, \K)$:
  $$\dom(\omega) \in \adm(q)$$
\end{requirement}

This constraint on variable bindings is still arguably weak though,
if one consider queries with \UNION.
Take for instance the query $q = A(x)\ \UNION\ R(x,y)$.
Then $\adm(q) = \{\{x\}, \{x,y\}\}$.
But the semantics of \sparql over plain graphs puts a stronger requirement on variable bindings.
If $\omega$ is a solution to $q$,
then $\omega$ may bind $\{x\}$ only if $\omega$ is an answer to the left operand $A(x)$,
and $\omega$ may bind $\{x,y\}$ only if $\omega$ is an answer to the right operand $R(x,y)$.
It is immediate to see that Requirement~\ref{req:var_bind} on variable bindings does not enforce this property.
So we add as a simple fifth requirement:
\begin{requirement}[Binding provenance]\label{req:var_bind_prov}
  For any SUJO queries $q_1, q_2$, KB $\K$ and solution mapping $\omega$:
  \begin{align*}
  \text{if } \omega \in \ans(q_1\ \UNION\ q_2, \K) \text{ and } \omega \not\in \ans(q_2), \text{ then }\dom(\omega) \in \adm(q_1)\\
  \text{if } \omega \in \ans(q_1\ \UNION\ q_2, \K) \text{ and } \omega \not\in \ans(q_1), \text{ then }\dom(\omega) \in \adm(q_2)
  \end{align*}
\end{requirement}


\section{Semantics}
\label{sec:semantics}
We now investigate different semantics for answering \sparql queries over a KB,
in view of the requirements expressed in the previous section. 
We note that each semantics is defined for a specific fragment of \sparql only,
and that this is also the case of Requirements~\ref{req:ca_comp}, \ref{req:var_bind} and~\ref{req:var_bind_prov} (the other two requirements are defined for arbitrary \sparql queries). 
So when we say below that a semantics defined for fragment $L_1$ \emph{satisfies} a requirement defined for fragment $L_2$,
this means that the requirement holds for the fragment $L_1 \cap L_2$.

Section~\ref{sec:compositional} shows that adopting a compositional interpretation or certain answers,
analogous to \sparql entailment regimes (restricted to SUJO queries),
is sufficient to satisfy Requirement~\ref{req:sa_comp},
but fails to satisfy Requirement~\ref{req:ca_comp} for the SJ and U fragments already.
Section~\ref{sec:canAns} focuses on DLs with the canonical model property.
For these, we consider generalizing a well-known property of certain answers to UCQs:
they are equivalent to answers over the canonical model,
but restricted to those that range over the active domain of the KB.
We show that this solution satisfies Requirements~\ref{req:ca_comp} and~\ref{req:sa_comp} for the SUJO fragment,
but fails to satisfy Requirement~\ref{req:op_mon} for the O fragment already.
Finally, Section~\ref{sec:certCanAns} builds upon this last observation,
and shows that it is possible to define a semantics that satisfies all requirements for the SUJO fragment.

Table~\ref{tab:semantics} summarizes our observations (for KBs with the canonical model property only),
together with observations about the proposal made in~\cite{ahmetaj_towards_2015} (discussed in Section~\ref{sec:sparql-over-kb}).

\begin{table}[tb]
  \centering
  \begin{tabular}{llccccc}
    \multicolumn{1}{l}{{\textbf{Semantics}}}
    & \multicolumn{1}{l}{\textbf{Fragment}}
    &\multicolumn{1}{c}{\textsc{req}\ref{req:ca_comp}}
    & \multicolumn{1}{c}{\textsc{req}\ref{req:sa_comp}}
    & \multicolumn{1}{c}{\textsc{req}\ref{req:op_mon}}
    & \multicolumn{1}{c}{\textsc{req}\ref{req:var_bind}}
    & \multicolumn{1}{c}{\textsc{req}\ref{req:var_bind_prov}}\\
                                                                                                                                            
    \toprule
    Ahmetaj et al.(\cite{ahmetaj_towards_2015})  &   pwdPT ($\subseteq$ SJO) & \checkmark & x & ? & \checkmark& \checkmark   \\ \midrule
Entailment regime (Def~\ref{def:compCertAns})  & UJO & \checkmark  & \checkmark  & \checkmark & \checkmark& \checkmark\\ 
               & SJ / SUJO & x & \checkmark & \checkmark & \checkmark & \checkmark\\ \midrule
    Canonical (Def~\ref{def:canAnswers}) & O / SUJO & \checkmark & \checkmark & x & \checkmark & \checkmark\\ \midrule
    Restricted (Def~\ref{def:restCanAns}) & SUJO & \checkmark & \checkmark& \checkmark& x & x   \\ \midrule
    Max. adm. can. (Def~\ref{def:certCanAns}) & SUJO & \checkmark & \checkmark& \checkmark& \checkmark & \checkmark \\ 
    \bottomrule
  \end{tabular}
  \vspace{2mm}
  \caption{Requirements met by alternative semantics for \sparql over a DL KB (with the canonical
    model property). ``A/B'' stands for all fragments between A and B.}
  \label{tab:semantics}
\end{table}

\subsection{\sparql Entailment Regimes}
\label{sec:compositional}
Example~\ref{ex:opt_ca} above showed that certain answer to a query with \OPT may fail to comply with the standard compositional semantics of \sparql (Definition~\ref{def:sparqlAns}) over a plain graph (i.e. when the TBox is empty).
Then a natural attempt to conciliate the two is to proceed ``the other way around'':
stick to the compositional semantics of \sparql,
and use certain answers for the base case only.
This is in essence what the \sparql entailment regimes propose for queries that correspond to the SUJO fragment (recall the restrictions
on reserved \rdf/\rdfs/\owl keywords in triple patterns expressed in Section~\ref{sec:prelim}).

Because the specification of \sparql entailment regimes~\cite{sparql-regimes} is too low-level for the scope of this paper, 
we provide a more abstract characterization of this approach for the SUJO fragment.
If $q$ is a query and $\K$ a KB,
we call the resulting set of solution mapping the \emph{entailment regime answers} to $q$ over $\K$,
denoted with $\cCertAns(q,\K)$,
defined as follows:
 \begin{definition}[Entailment Regime Answers]\label{def:compCertAns}
\begin{center}
  \begin{tabular}{lll}
    \multicolumn{3}{l}{If $q$ is a triple pattern, then $\cCertAns(q,\K) = \certAns(q,\K)$}\\
    $\cCertAns(q_1\ \UNION\ q_2,  \K)$& $=$& $\cCertAns(q_1, \K) \cup \cCertAns(q_2, \K)$\\
    $\cCertAns(q_1\ \JOIN\ q_2, \K)$& $=$& $\cCertAns(q_1, \K) \bowtie \cCertAns(q_2, \K)$\\
    $\cCertAns(q_1\ \OPT\ q_2, \K)$& $=$& $(\cCertAns(q_1, \K) \bowtie \cCertAns(q_2,\K))\ \cup$\\
                                      &&$(\cCertAns(q_1,\K) \setminus \cCertAns(q_2,\K))$\\
    $\cCertAns(\SELECT_X\ q, \K)$ &$=$&  $\pi_X \cCertAns(q, \K)$
\end{tabular}
    \end{center}
 \end{definition}
 It is immediate to see that entailment regime answers and \sparql answers coincide over a plain graph.
 Indeed,
 in the base case (i.e. when $q$ is a triple pattern), for any graph $\A$,
 $\sparqlAns(q,\A) = \certAns(q,\tup{\emptyset, \A})$.
 Then the inductive definitions of $\sparqlAns(q,\A)$ (Definition~\ref{def:sparqlAns}) and $\cCertAns(q, \K)$ (Definition~\ref{def:compCertAns}) coincide.
 So entailment regime answers satisfy Requirement~\ref{req:sa_comp}.

 But they fail to comply with certain answers for UCQs (Requirement~\ref{req:ca_comp}),
 for two reasons.
 First, the \UNION operator is not compositional for certain answers in some DLs.
 Consider for instance Example~\ref{ex:nonCompUnion} below:
 \begin{example}\label{ex:nonCompUnion}\ \\
   $\A = \{\stt{Driver(Alice)}\}$\\
   $\T = \{\stt{Driver} \sqsubseteq \stt{CarDriver} \sqcup \stt{TruckDriver}\}$\\
   $q\ = \stt{CarDriver}(x)\ \UNION\ \stt{TruckDriver}(x)$

   \noindent   
 Then $\certAns(q,\tup{\T,\A}) = \{\{x \mapsto \stt{Alice}\}\}$,
 but $\cCertAns(q,\tup{\T,\A}) = \emptyset$.
 \end{example}

Second,
the \SELECT operator is not compositional for certain answers,
even for some DLs that have the canonical model property.
Consider for instance Example~\ref{ex:nonComp} below:
\begin{example}\label{ex:nonComp}\ \\
  $\A = \{\stt{Driver(Alice)}\}$\\
  $\T = \{\stt{Driver} \sqsubseteq \exists\stt{hasLicense}\}$\\
  $q\ = \SELECT_{\{x\}}\ (\stt{Driver}(x)\ \JOIN\
  \stt{hasLicense}(x,y))$

  \noindent
  Then $\certAns(q,\tup{\T,\A}) = \{\{x \mapsto \stt{Alice}\}\}$, but
  $\cCertAns(q, \tup{\T,\A}) = \emptyset$.
 \end{example}

 So entailment regime answers fail to satisfy Requirement~\ref{req:ca_comp} for the U and SJ fragments already.

\subsection{Canonical Answers}
\label{sec:canAns}
We now focus on DLs with the canonical model property.
We assume some underspecified DL $\lcan$ with the canonical model property,
and use ``an $\lcan$ KB'' to refer to a KB in such DL.
Then if $\K$ is an $\lcan$ KB,
we use $\can(\K)$ to designate its canonical model (up to isomorphism).

An equivalent definition of certain answers
for DLs with the canonical model property is the following:
certain answers to a UCQ $q$ over a KB $\K$ coincide with answers to $q$ over $\can(\K)$,
restricted to those that range over $\aDom(\K)$.
We show that extending this definition to queries with \OPT is sufficient to satisfy Requirements~\ref{req:sa_comp} (in addition to Requirement~\ref{req:ca_comp}), 
but fails to satisfy Requirement~\ref{req:op_mon}.

If $\Omega$ is a set of solution mappings and $B \subseteq \NI$,
let $\Omega \vartriangleright B = \{\omega \in \Omega \mid \range(\omega) \subseteq B\}$.
Then we define the \emph{canonical answers} to a query $q$ over an $\lcan$ KB $\K$,
denoted with $\canAns(q,\K)$,
as follows:
\begin{definition}[Canonical Answers]\label{def:canAnswers}
  For any SUJO query $q$ and $\lcan$ KB $\K$:
$$\canAns(q,\K) = \restRangeIn{\sparqlAns(q,\can(\K))}{\aDom(\K)}$$
\end{definition}

Proposition~\ref{prop:cana_compliance} states that canonical answers comply with \sparql answers over a plain graph (Requirement~\ref{req:sa_comp}).

\begin{restatable}{proposition}{propCanaCompliance}\label{prop:cana_compliance}
  For any SUJO query $q$ and $\lcan$ KB $\K$,
  $\canAns(q,\K)$ satisfies Requirement~\ref{req:sa_comp}.
\end{restatable}
From the observation made above,
canonical answers also comply with certain answers for UCQs (Requirement~\ref{req:ca_comp}).
But they fail to satisfy \OPT extension (Requirement~\ref{req:op_mon}),
as illustrated with Example~\ref{ex:canAns_opt}.
 \begin{example}\label{ex:canAns_opt}\ \\
   $\A = \{\stt{Driver(Alice)}\}$\\
   $\T = \{\stt{Driver} \sqsubseteq \exists\stt{hasLicense}\}$\\
   $q\ = \stt{Driver}(x)\ \OPT\ \stt{hasLicense}(x,y)$
 \end{example}
 In this example,
 Let $\K = \tup{\T,\A}$.
 Then $\canAns(\stt{Driver}(x),\K) = \{\{ x \mapsto \stt{Alice}\}\}$.
 However, $\sparqlAns(q, \can(\K)) = {\{\{ x \mapsto \stt{Alice}, y \mapsto e\}\}}$,
 for some $e \not\in \aDom(\K)$.
 Therefore $\canAns(q,\K) = \restRangeIn{\sparqlAns(q,\can(\K))}{\aDom(\K)} = \emptyset$.
 So $\canAns(\stt{Driver}(x),\K) \not\preceq_g \canAns(q,\K)$,
 which immediately violates Requirement~\ref{req:op_mon}.

 \subsection{Maximal Admissible Canonical Answers}
 \label{sec:certCanAns}
 The canonical answers defined in the previous section fail to satisfy Requirement~\ref{req:op_mon}.
We show how this definition can be adapted to satisfy all requirements,
for the whole SUJO fragment.

Intuitively,
in Definition~\ref{def:canAnswers},
the restriction of $\sparqlAns(q,\can(\K))$ to solution mappings that range over $\can(\K)$ is too strong.
Consider again Example~\ref{ex:canAns_opt},
where $\sparqlAns(q, \can(\K)) = \{\{ x \mapsto \stt{Alice}, y \mapsto e\}\}$.
In this example,
rather than filtering out this solution mapping (because it does not range over $\aDom(\K)$),
one would want instead to \emph{restrict} it to the active domain,
which yields the desired mapping $\{x \mapsto \stt{Alice}\}$.

To formalize this intuition,
if $\Omega$ is a set of solution mappings and $B \subseteq \NI$,
let $\Omega \blacktriangleright B = \{\omega\|_B \mid \omega \in \Omega\}$.
We can now define the \emph{restricted canonical answers} $\restCanAns(q,\K)$ to a query $q$ over an $\lcan$ KB $\K$,
as follows:
\begin{definition}[Restricted Canonical Answers]\label{def:restCanAns}
  For any SUJO query $q$ and $\lcan$ KB $\K$: $$\restCanAns(q,\K) = \restRange{\sparqlAns(q,\can(\K))}{\aDom(\K)}$$
\end{definition}

However,
restricted canonical answers still fail to satisfy the above requirement on admissible variable bindings (Requirement~\ref{req:var_bind}),
as illustrated with Example~\ref{ex:rest_failure} below:
 \begin{example}\label{ex:rest_failure}\ \\
   $\A = \{\stt{Teacher(Alice)}\}$\\
   $\T = \{\stt{Teacher} \sqsubseteq \exists \stt{teachesTo}, \stt{teachesTo} \sqsubseteq \stt{hasTeacher}^-\}$\\
   $q\ = \stt{Teacher}(x)\ \OPT\ (\stt{teachesTo}(x,y)\ \JOIN\ \stt{hasTeacher}(y,z))$
 \end{example}
 In this example,
 $\sparqlAns(q,\can(\K)) = \{\{ x \mapsto \stt{Alice}, y \mapsto e, z \mapsto \stt{Alice}\}\}$,
 for some $e \not\in \aDom(\K)$.
 So restricting this solution mapping to $\aDom(\K)$ would yield the mapping $\{ x \mapsto \stt{Alice}, z \mapsto \stt{Alice}\}$.
 However, $\{x,z\}$ is not an admissible set of variables for $q$,
 because $q$ requires that whenever variable $z$ is bound,
 variable $y$ must be bound as well.

 We now propose to further constrain restricted canonical answers in order to satisfy Requirements~\ref{req:var_bind} and~\ref{req:var_bind_prov}.
 We call the resulting solution mappings \emph{maximal admissible canonical answers},
 noted $\certCanAns(q,\K)$.

 We start with the PJO fragment (i.e. queries without \UNION) for simplicity,
 since for this fragment,
 Requirement~\ref{req:var_bind_prov} is trivially satisfied. 
  If $\S$ is a family of sets, let $\max_\subseteq (\S)$ designate the set of maximal elements of $\S$ w.r.t. set inclusion.
 And if $\Omega$ is a set of solution mappings and $\X$ a family of sets of variables,
let:
$$\textstyle\Omega \otimes \X =\ \{\omega|_X \mid \omega\in\Omega, X \in \max_\subseteq (\X \cap 2^{\dom(\omega)})\}$$

We can now define maximal admissible canonical answers for the SJO fragment,
as follows:
\begin{definition}[Maximal Admissible Canonical Answers (SJO)]\label{def:certCanAns_SJO}\ \\
$$\certCanAns(q, \K) = \restCanAns(q, \K)  \otimes \adm(q)$$
\end{definition}
In order to generalize this definition to queries with \UNION,
we need to enforce Requirement~\ref{req:var_bind_prov}.
To this end, the provenance of each solution mapping needs to be taken into account.
We define the set of \emph{branches} of a SUJO query $q$,
denoted with $\branch(q)$,
as the set of SJO queries that may produce a solution to $q$,
by intuitively ``choosing'' one operand of each \UNION.
For instance,
if $q = A(x)\ \OPT\ (R_1(x,y)\ \UNION\ R_2(x,z))$,
then $\branch(q) = \{A(x)\ \OPT\ R_1(x,y), A(x)\ \OPT\ R_2(x,z)\}$.
The function $\branch(q)$ is defined inductively over $q$ as expected:
\clearpage
\begin{definition}[Branches of a SUJO query $q$]\label{def:branch}
\begin{center}
  \begin{tabular}{lll}
    \multicolumn{3}{l}{If $q$ is a triple pattern, then $\branch(q) = \{q\}$}\\
    $\branch(\SELECT_X\ q)$ &$= \{\ \SELECT_X\ q'$&$\mid q' \in \branch(q)\ \}$\\
    $\branch(q_1\ \JOIN\ q_2)$ &$= \{\ q'_1\ \JOIN\ q'_2$&$\mid (q'_1,q'_2) \in \branch(q_1) \times \branch(q_2)\ \}$\\
    $\branch(q_1\ \OPT\ q_2)$ &$= \{\ q'_1\ \OPT\ q'_2$&$\mid (q'_1,q'_2) \in \branch(q_1) \times \branch(q_2)\ \}$\\
    $\branch(q_1\ \UNION\ q_2)$ &\multicolumn{2}{l}{$= \branch(q_1) \cup \branch(q_2)$}
\end{tabular}
\end{center}
\end{definition}

\noindent
According to the semantics of \sparql over plain graphs,
an answer to a SUJO query $q$ must be an answer to some branch of $q$ (the converse does not hold though; see e.g.~\cite[Example 1]{schmidt_foundations_2010}).
Or formally, for any SUJO query $q$ and graph $\A$:
$$\sparqlAns(q, \A) \subseteq \bigcup\limits_{q'\ \in \branch(q)} \sparqlAns(q', \A)$$
So if $q' \in \branch(q)$,
we use $\sparqlAns(q, \A, q')$ to denote the answers to $q$ over $\A$ that may be obtained by evaluating branch $q'$,
i.e.:
$$\sparqlAns(q, \A, q') = \sparqlAns(q, \A) \cap  \sparqlAns(q', \A)$$
Similarly, we adapt Definition~\ref{def:certCanAns_SJO} to a branch $q'$ of $q$:
$$\certCanAns(q, \K, q') = (\restRange{\sparqlAns(q, \can(\K), q')}{\aDom(\K)}) \otimes \adm(q')$$

\noindent
We can now generalize maximal admissible canonical answers to the SUJO fragment:
\begin{definition}[Maximal Admissible Canonical Answers (SUJO)]\label{def:certCanAns}
$$\certCanAns(q, \K) = \bigcup\limits_{q' \in \branch(q)} \certCanAns(q, \K, q')$$
\end{definition}
It can be easily verified that Definitions~\ref{def:certCanAns_SJO} and~\ref{def:certCanAns} coincide for SJO queries,
since in this case $\branch(q) = \{q\}$.
Proposition~\ref{prop:ecanAns_comp} shows that maximal admissible canonical answers satisfy all requirements expressed in the previous section.

\begin{restatable}{proposition}{propEcanAnsComp}\label{prop:ecanAns_comp}
  For any SUJO query $q$ and $\lcan$ KB $\K$,
  $\certCanAns(q,\K)$ satisfies Requirements~\ref{req:ca_comp}, \ref{req:sa_comp}, \ref{req:op_mon}, \ref{req:var_bind} and~\ref{req:var_bind_prov}.
\end{restatable}





\section{Complexity}
\label{sec:complexity}

We now provide complexity results for query answering under the semantics defined in Section~\ref{sec:certCanAns},
for different sub-fragments of the SUJO fragment,
and focusing on KBs in \dlliter~\cite{artale2009dl},
a DL tailored for query answering,
which corresponds to the \owltwoql profile.
As is conventional,
we focus on the \emph{decision problem} for query answering,
i.e. the problem \textsc{eval}$_{\certCanAns}$ below.
We also focus on \emph{combined} complexity, i.e. measured in the size of the whole input (KB and query),
and leave \emph{data} complexity (parameterized either by the size of the query, or of the query and TBox) as future work.

\begin{center}
 \fbox{
 \begin{tabular}{l}
 \textsc{eval}$_{\certCanAns}$ \\
 \begin{tabular}{ll}
   \textbf{Input}: & \! \dlliter KB $\K$, query $q$, mapping $\omega$ \\
 \textbf{Decide}: & $\omega \in \certCanAns(q, \K)$
 \end{tabular} \\
 \end{tabular}
 }
 \end{center}
 Complexity of \sparql query evaluation over plain graphs has been extensively studied (see~\cite{mengel_tractable_2017} for a recent overview).
When these results are tight, they provide us immediate lower bounds.
Indeed, from Proposition~\ref{prop:cana_compliance},
certain canonical answers satisfy Requirement~\ref{req:sa_comp},
so \textsc{eval}$_{\certCanAns}$ is at least as hard as the problem \textsc{eval}$_{\sparqlAns}$ below.
 \begin{center}
 \fbox{
 \begin{tabular}{l}
 \textsc{eval}$_{\sparqlAns}$ \\
 \begin{tabular}{ll}
   \textbf{Input}: & \! graph $\A$, query $q$, mapping $\omega$ \\
 \textbf{Decide}: & $\omega \in \sparqlAns(q, \A)$
 \end{tabular} \\
 \end{tabular}
 }
\end{center}
Table~\ref{tab:complexity} reproduces results for \textsc{eval}$_{\sparqlAns}$ in several commonly studied fragments that fall within the SUJO fragment.
 The \OPT operator has been the focus of a large part of the literature,
 as \textsc{eval}$_{\sparqlAns}$ has been shown to be \PSPACE-complete for the OJ fragment already,
 in~\cite{schmidt_foundations_2010}.
 Particular attention has also been paid to so-called \emph{well-designed} SJO and JO queries (see~\cite{perez_semantics_2009}  for a definition),
 which have a natural representation as \emph{pattern trees}~\cite{letelier_static_2013},
 with a significant reduction from \PSPACE to \sigmaptwo and \coNP-completeness respectively.
 For SJO, we follow~\cite{letelier_static_2013} and focus on queries where the \SELECT operator is terminal,
 i.e. where it does not appear in the scope of \JOIN or \OPT.
 The corresponding fragment is called SJO*.
 Finally, another fragment of interest is UJ, for which query answering is already intractable~\cite{schmidt_foundations_2010},
 thus contrasting with projection-free UCQs.

 So for each fragment,
 we investigate whether \textsc{eval}$_{\certCanAns}$ matches the upper bounds for \textsc{eval}$_{\sparqlAns}$.
 The results are summarized in Table~\ref{tab:complexity}.
\begin{table}[tb]
  \centering
  \begin{tabular}{lll}
    \multicolumn{1}{l}{\textbf{Fragment}}
    &\multicolumn{1}{c}{\textsc{eval}$_{\sparqlAns}$}
    & \multicolumn{1}{c}{\textsc{eval}$_{\certCanAns}$}\\
    \toprule
    UJ/SUJ & \NP-c& \NP-c\\
    well-designed JO & \coNP-c& \coNP-c\\
    well-designed SJO* & \sigmaptwo-c & \sigmaptwo-c\\
    OJ/SUJO & \PSPACE-c& \PSPACE-c\\
    \bottomrule
  \end{tabular}
  \vspace{2mm}
  \caption{Combined complexity of \textsc{eval}$_{\sparqlAns}$ and \textsc{eval}$_{\certCanAns}$.\\
    ``-c'' stands for complete, and
    ``A/B'' for all fragments between A and B.}
  \label{tab:complexity}
\end{table}
Interestingly,
all upper bounds are matched.
This means that for these fragments,
the presence of a TBox does not induce an extra computational cost (as far as worst-case complexity is concerned) when compared to \sparql answers over a plain graph.
This observation is analogous to well-known results for answering UCQs under certain-answer semantics over a \dlliter KB~\cite{calvanese2007tractable},
which matches the (\NP) upper bound for UCQs over a plain graph. 

Before explaining these results,
we isolate a key observation:
 \begin{restatable}{proposition}{propCostAdm}\label{prop:cost_adm}
   If $q$ is a JO query and $X_1,X_2 \subseteq \vars(q)$, then it can be decided in $O(|q|^2)$ whether $X_1 \in \max_\subseteq(\adm(q) \cap 2^{X_2})$
 \end{restatable}
 \begin{proof} (Sketch.)
   If $q$ is a JO query, we compute a family $\base(q)$ of sets of variables s.t. $|\base(q)| = O(|q|)$,
   and s.t. each $V \in \adm(q)$ is the union of some elements of $\base(q)$ and conversely,
   i.e. $\adm(q) = \{\bigcup \B \mid \B \in 2^{\base(q)}\}$.
   The family $\base(q)$ can be computed inductively as follows:
   \begin{itemize}
   \item if $q$ is a triple pattern, then $\base(q) = \{\vars(q)\}$.
   \item if $q = q_1\ \JOIN\ q_2$, then $\base(q) = \{B_1 \cup B_2 \mid B_1 \in \min_\subseteq (\base(q_1)), B_2 \in \base(q_2)\} \cup$\\
     $\{B_1 \cup B_2 \mid B_1 \in \base(q_1), B_2 \in \min_\subseteq (\base(q_2))\}$
   \item if $q = q_1\ \OPT\ q_2$, then $\base(q) = \base(q_1) \cup \base(q_1\ \JOIN\ q_2)$
   \end{itemize}
   The induction guarantees that $|\min_\subseteq (\base(q))| = 1$, so that $|\base(q))| = O(|q|)$ must hold.
   Then in order to decide $X_1 \in \max_\subseteq(\adm(q) \cap 2^{X_2})$, it is sufficient to:
   \ei check whether $X_1 \in \adm(q)$, i.e. check whether $X_1 \subseteq \bigcup\{B \in \base(q) \mid B \subseteq X_1\}$,
   and
   \eii check whether there is an $X' \in \adm(q) \cap 2^{X_2}$ s.t. $X \subsetneq X'$.
   This is the case iff there is a $B \in \base(q)$ s.t. $X_1 \subsetneq X_1 {\cup} B \subsetneq X_2$.
 \end{proof}
 We note that from the definition of $\adm(q)$,
 this property is independent from the semantics under investigation,
 so it holds for \sparql over a plain graph.
 It also follows that deciding whether $X \in \adm(q)$ for an arbitrary $X$ and JO query $q$ is tractable (consider the case where $X_1 = X_2$).
 Interestingly, this does not hold for the UJ fragment already.
 Indeed,
 immediately from the reduction used in~\cite{schmidt_foundations_2010} for hardness of \textsc{eval}$_{\sparqlAns}$ in this fragment,
 deciding $X \in \adm(q)$ for any $X$ and UJ query $q$ is \NP-hard (see the appendix for details).

 We now sketch the argument used to derive upper bounds for the SUJO, well-designed SJO* and UJ fragments (proofs can be found in the appendix).
 For simplicity, we focus on the well-designed SJO* fragment.
 The argument for queries with \UNION is similar, but with additional technicalities,
 because the definition of maximal admissible canonical answers in this case is more involved (compare Definitions~\ref{def:certCanAns_SJO} and~\ref{def:certCanAns} above).
 We also simplify the explanation by assuming that the Gaifman graph of the query is connected.
 If $\G$ is a graph, we will use $V(\G)$ below to designate its vertices.
 
 From the definition of \textsc{eval}$_{\certCanAns}$,
 $\tup{\K ,q, \omega}$ is a positive instance iff $\omega \in \certCanAns(q, \K)$,
 i.e. iff there is an $\omega'$ s.t.
\ei $\omega = \omega'|_X$ for some $X \in \max_\subseteq (\adm(q) \cap 2^{\dom(\omega'\|_{\aDom(\K)})})\}$ and \eii $\omega' \in \sparqlAns(q,\K)$.

So a (non-deterministic) procedure to verify $\omega \in \certCanAns(q, \K)$ consists in guessing an extension $\omega'$ or $\omega$,
then verify \ei, and then verify \eii.
From Proposition~\ref{prop:cost_adm} above,
\ei can be verified in $O(|q|^2)$.
For \eii,
if $\omega' \in \sparqlAns(q,\can(\K))$,
from well-known properties of $\can(\K)$ for $\dlliter$,
it can be shown that:
\begin{itemize}
\item there must exist a subgraph $\G$ of $\can(\K)$ s.t. $V(\G) \cap V(\A) \neq \emptyset$,
  and the size of the subgraph of $\G$ induced by $V(\G) \setminus V(\A)$ is linearly bounded by $\max(|q|, |\T|)$.
\item 
for each maximal connected subgraph $\G'$ of $\G$ s.t. $V(\G') \cap V(\A) = \emptyset$,
it can be verified in $O((|\G'|+|\T|) \cdot |\T|)$ whether $\G'$ is a subgraph of $\can(\K)$.
\end{itemize}
So in order to verify \eii,
it is sufficient to guess $\G$,
then verify that $\G$ is a subgraph of $\can(\K)$,
and then verify that $\omega' \in \sparqlAns(q,\G)$.
Since \textsc{eval}$_{\sparqlAns}$ is in \sigmaptwo,
$\omega' \in \sparqlAns(q,\G)$ can be nondeterministically verified in time in $O(|q|+|\G|+|\omega'|) = O(|q|+|\K|+\omega)$ by some algorithm with an oracle for $\coNP$ problems.
And a witness for this algorithm can be guessed together with $\G$ and $\omega'$ (without gaining a level in the polynomial hierarchy).
We note that this last remark does not apply to the well-designed JO fragment:
since \textsc{eval}$_{\sparqlAns}$ is $\coNP$-hard,
such a procedure would instead imply a quantifier alternation.

The proof of $\coNP$-membership for the well-designed JO fragment is significantly simpler.
First, because the fragment does not allow projection,
for any JO query $q$,
$\certCanAns(q,\K) = \canAns(q,\K)$ must hold.
Then we consider the ABox $\A'$ that contains all atoms over the active domain that are entailed by $\K$,
i.e. $\A' = \{ A(c) \in \can(\K) \mid c \in \aDom(\K) \} \cup \{ r(c_1, c_2) \in \can(\K) \mid c_1,c_2 \in \aDom(\K) \}$.
$\A'$ can be computed in time polynomial in $\K$ and,
by immediate induction on $q$,
it can be shown that $\canAns(q,\K) = \sparqlAns(q,\A')$.
Finally,
from~\cite{perez_semantics_2009},
$\omega \in \sparqlAns(q,\A')$ is in $\coNP$.


\clearpage
\section{Conclusion and Perspectives}
\label{sec:conclusion}

We identified in this article simple properties to be met by a semantics meant to conciliate certain answers to UCQs over a KB on the one hand, and \sparql answers over a plain graph on the other hand.
We formalized these properties as requirements,
and evaluated different proposals (some of which were taken from the literature) against these requirements.

We also showed that these requirements can be all satisfied for the fragment of \sparql with \stt{SELECT}, \stt{UNION} and \stt{OPTIONAL} and DLs with the canonical model property.
More precisely,
we defined a semantics that matches all requirements.
We also provided combined complexity results for query answering over a \dlliter KB under this semantics.

This work is still at an early stage,
for multiple reasons.
First,
the semantics we defined is arguably ad-hoc,
with a procedural flavor,
and it would be interesting to investigate whether it can be characterized in a more declarative fashion.
It must also be emphasized that if query answers defined by this semantics comply with all requirements,
whether the converse holds (i.e. whether there may be answers that comply with all requirements,
but are not returned under this semantics) is still an open question.

Data complexity may also be investigated,
as well as algorithmic aspects,
in particular \emph{FO-rewritability},
i.e. the possibility to rewrite a query over a KB into a query over its ABox only,
which is a key property for OMQA/OBDA~\cite{2018-ijcai-obda-survey}.
Other DLs and/or fragments of \sparql may also be considered. 


Finally, and more importantly,
additional requirements may be identified,
possibly violated by the semantics we defined.
If so,
a key question is whether such an extended set of requirements can still be matched,
for reasonably expressive DLs and fragments of \sparql.
A negative answer would constitute an argument for the \sparql entailment regimes (or the extension of the \owltwoql regime proposed in~\cite{kontchakov_answering_2014}) as a default solution.






  \newpage
{\small
\bibliography{bib} 

\begin{thebibliography}{10}

\bibitem{ahmetaj_towards_2015}
S.~Ahmetaj, W.~Fischl, R.~Pichler, M.~Šimkus, and S.~Skritek.
\newblock Towards reconciling {SPARQL} and certain answers.
\newblock In {\em Proceedings of the 24th {International} {Conference} on
  {World} {Wide} {Web}}, pages 23--33. ACM, 2015.

\bibitem{arenas_querying_2011}
M.~Arenas and J.~Pérez.
\newblock Querying semantic web data with {SPARQL}.
\newblock In {\em Proceedings of the thirtieth {ACM} {SIGMOD}-{SIGACT}-{SIGART}
  symposium on {Principles} of database systems}, pages 305--316. ACM, 2011.

\bibitem{artale2009dl}
A.~Artale, D.~Calvanese, R.~Kontchakov, and M.~Zakharyaschev.
\newblock The dl-lite family and relations.
\newblock {\em Journal of artificial intelligence research}, 36:1--69, 2009.

\bibitem{dl-handbook}
F.~Baader, D.~Calvanese, D.~McGuinness, D.~Nardi, and P.~Patel-Schneider,
  editors.
\newblock {\em The Description Logic Handbook: Theory, Implementation, and
  Applications}.
\newblock Cambridge University Press, 2003.

\bibitem{calvanese2007tractable}
D.~Calvanese, G.~De~Giacomo, D.~Lembo, M.~Lenzerini, and R.~Rosati.
\newblock Tractable reasoning and efficient query answering in description
  logics: The dl-lite family.
\newblock {\em Journal of Automated reasoning}, 39(3):385--429, 2007.

\bibitem{sparql-regimes}
B.~Glimm and C.~Ogbuji.
\newblock {SPARQL} 1.1 entailment regimes.
\newblock Technical report, W3C, March 2013.

\bibitem{gutierrez2016certain}
C.~Gutierrez, D.~Hern{\'a}ndez, A.~Hogan, and A.~Polleres.
\newblock Certain answers for sparql?
\newblock In {\em AMW}, 2016.

\bibitem{sparql11}
S.~Harris, A.~Seaborne, and E.~Prud'hommeaux.
\newblock {SPARQL} 1.1 query language.
\newblock {W3C} recommendation, {W3C}, 2013.

\bibitem{hernandez2018certain}
D.~Hern{\'a}ndez, C.~Gutierrez, and A.~Hogan.
\newblock Certain answers for sparql with blank nodes.
\newblock In {\em International Semantic Web Conference}, pages 337--353.
  Springer, 2018.

\bibitem{kontchakov_answering_2014}
R.~Kontchakov, M.~Rezk, M.~Rodríguez-Muro, G.~Xiao, and M.~Zakharyaschev.
\newblock Answering {SPARQL} queries over databases under {OWL} 2 {QL}
  entailment regime.
\newblock In {\em International {Semantic} {Web} {Conference}}, pages 552--567.
  Springer, 2014.

\bibitem{kostylev_semantics_2014}
E.~V. Kostylev and B.~C. Grau.
\newblock On the semantics of {SPARQL} queries with optional matching under
  entailment regimes.
\newblock In {\em International {Semantic} {Web} {Conference}}, pages 374--389.
  Springer, 2014.

\bibitem{letelier_static_2013}
A.~Letelier, J.~Pérez, R.~Pichler, and S.~Skritek.
\newblock Static analysis and optimization of semantic web queries.
\newblock {\em ACM Transactions on Database Systems (TODS)}, 38(4):25, 2013.

\bibitem{mengel_tractable_2017}
S.~Mengel and S.~Skritek.
\newblock On tractable query evaluation for {SPARQL}.
\newblock {\em arXiv preprint arXiv:1712.08939}, 2017.

\bibitem{perez_semantics_2009}
J.~Pérez, M.~Arenas, and C.~Gutierrez.
\newblock Semantics and complexity of {SPARQL}.
\newblock {\em ACM Transactions on Database Systems (TODS)}, 34(3):16, 2009.

\bibitem{schmidt_foundations_2010}
M.~Schmidt, M.~Meier, and G.~Lausen.
\newblock Foundations of {SPARQL} query optimization.
\newblock In {\em Proceedings of the 13th {International} {Conference} on
  {Database} {Theory}}, pages 4--33. ACM, 2010.

\bibitem{2018-ijcai-obda-survey}
G.~Xiao, D.~Calvanese, R.~Kontchakov, D.~Lembo, A.~Poggi, R.~Rosati, and
  M.~Zakharyaschev.
\newblock Ontology-based data access: A survey.
\newblock In {\em Proceedings of the Twenty-Sixth International Joint
  Conference on Artificial Intelligence, {IJCAI-18}}, pages 5511--5519.
  International Joint Conferences on Artificial Intelligence Organization, 7
  2018.

\end{thebibliography}
\bibliographystyle{abbrv} 
}

  \newpage
  \appendix
 \section{Proof of Proposition~\ref{prop:cana_compliance}}
\propCanaCompliance*

\begin{proof}\ \\
  Lemma~\ref{lemma:canAns_sparqlComp} below states the proposition.
\end{proof}

\begin{lemma}\label{lemma:canAns_sparqlComp}
  For any UCQ $q$ and ABox $\A$,
  $$\canAns(q,\tup{\emptyset,\A}) = \sparqlAns(q,\A)$$
\end{lemma}
\begin{proof}\ \\
  if $\A$ is an ABox,
  then $\aDom(\tup{\A,\emptyset})$ is the set of constants appearing in $\A$.\\
  In addition,
  $\can(\tup{\A,\emptyset}) = \A$.\\
  So if $q$ is a query, trivially,
  $\restRangeIn{\sparqlAns(q,\tup{\A,\emptyset})}{\aDom(\tup{\A,\emptyset})} = \sparqlAns(q,\tup{\A,\emptyset})$.\\
  So from Definition\ref{def:canAnswers},
  $\canAns(q,\tup{\A,\emptyset}) = \sparqlAns(q,\tup{\A,\emptyset})$.
\end{proof}

\section{Proof of Proposition~\ref{prop:ecanAns_comp}}
\propEcanAnsComp*
\begin{proof}\ \\
  The proposition is split into Lemmas~\ref{lemma:certCanAns_caComp}, \ref{lemma:certCanAns_saComp}, 
  \ref{lemma:certCanAns_optComp}, \ref{lemma:certCanAns_vbComp} and~\ref{lemma:certCanAns_provComp} below,
  one for each requirement.
\end{proof}

\begin{lemma}\label{lemma:certCanAns_caComp}
  For any UCQ $q$ and $\lcan$ KB $\K$,
  $\certCanAns(q,\K) = \certAns(q,\K)$
\end{lemma}
\begin{proof}
  Let $q$ be a UCQ and $\K$ an $\lcan$ KB.\\
  Lemma~\ref{lemma:certCanAns_canAns_corresp} below states that $\certCanAns(q,\K) = \canAns(q,\K)$.\\
  Then the claim follows immediately from the observation (made in~\ref{sec:canAns}) that $\canAns(q,\K)$ satisfies Requirement~\ref{req:ca_comp}.
\end{proof}

\begin{lemma}\label{lemma:certCanAns_canAns_corresp}
  For any UCQ $q$ and $\lcan$ KB $\K$,
  $\certCanAns(q,\K) = \canAns(q,\K)$
\end{lemma}
\begin{proof}
  Let $q$ be a UCQ.\\
  Then $q$ is of the form $h_1\ \UNION .. \UNION\ h_n$,
  where each $h_i$ can only contains \SELECT or \JOIN operators,
  and $\vars(h_i) = \vars(h_j)$ for all $i,j \in \{1..n\}$.\\
  So immediately from Definition~\ref{def:sparqlAns},
  for each $q' \in \branch(q)$,
  $\adm(q') = \{\vars(q)\}$.\\
  Therefore for any $q' \in \branch(q)$,
  $\adm(q') = \{\vars(q)\}$\\
  Then from the definition of $\certCanAns(q, \K, q')$:
  \begin{align}
\certCanAns(q, \K, q') &= (\restRange{\sparqlAns(q, \can(\K), q')}{\aDom(\K)}) \otimes \adm(q')\\
\certCanAns(q, \K, q') &= (\restRange{\sparqlAns(q, \can(\K), q')}{\aDom(\K)}) \otimes \{\vars(q)\}\\
\certCanAns(q, \K, q') &= \{\omega\|_{\aDom(\K)} \mid \omega \in \sparqlAns(q, \can(\K),q')\} \otimes \{\vars(q)\}
  \end{align}
Then for each $\omega \in \sparqlAns(q, \can(\K), q')$,
$\omega \in \sparqlAns(q, \can(\K))$ must hold.\\
So since $\adm(q) = \{\vars(q)\}$,
$\dom(\omega) = \vars(q)$ must hold as well.\\
Therefore $\vars(q) \subseteq \dom(\omega\|_{\aDom(\K)})$ iff
$\omega\|_{\aDom(\K)} = \omega$, i.e. iff $\range(\omega) \subseteq \aDom(\K)$.
So from~\ref{eq:5-0}:
\begin{align}
\certCanAns(q, \K, q') &= \{\omega \in \sparqlAns(q, \can(\K), q') \mid \range(\omega) \subseteq \aDom(\K)\}\label{eq:5-0}\\
\certCanAns(q, \K, q') &= \restRangeIn{\sparqlAns(q, \can(\K), q')}{\aDom(\K)}\label{eq:5-1}\\
\intertext{Finally:}
\certCanAns(q, \K) &= \bigcup\limits_{q' \in \branch(q)} \certCanAns(q, \K, q')\label{eq:5-2}\\
\intertext{So from~\ref{eq:5-1} and~\ref{eq:5-2}:}
\certCanAns(q, \K) &= \bigcup\limits_{q' \in \branch(q)} \restRangeIn{\sparqlAns(q, \can(\K), q')}{\aDom(\K)}\\
\certCanAns(q, \K) &= \bigcup\limits_{q' \in \branch(q)} \restRangeIn{(\sparqlAns(q, \can(\K)) \cap \sparqlAns(q', \can(\K)))}{\aDom(\K)}\\
\certCanAns(q, \K) &= \restRangeIn{\left(\sparqlAns(q, \can(\K)) \cap \bigcup\limits_{q' \in \branch(q)} \sparqlAns(q', \can(\K))\right)}{\aDom(\K)}
\intertext{And since:}
\sparqlAns(q, \K) &\subseteq \bigcup\limits_{q'\ \in \branch(q)} \sparqlAns(q', \can(\K))\\
\intertext{we get:}
\certCanAns(q, \K) &= \restRangeIn{\sparqlAns(q, \can(\K))}{\aDom(\K)}
\end{align}
\end{proof}

\begin{lemma}\label{lemma:certCanAns_saComp}
  For any SUJO query $q$ and ABox $\A$,
  $$\certCanAns(q,\tup{\emptyset,\A}) = \sparqlAns(q,\A)$$
\end{lemma}
\begin{proof}
  if $\A$ is an ABox,
  then $\aDom(\tup{\A,\emptyset})$ is the set of constants appearing in $\A$.\\
  In addition,
  $\can(\tup{\A,\emptyset}) = \A$.\\
  Then from the definition of $\certCanAns(q, \K, q')$:
  \begin{align}
\certCanAns(q, \tup{\A,\emptyset}, q') &= (\restRange{\sparqlAns(q, \A, q')}{\aDom(\tup{\A,\emptyset})}) \otimes \adm(q')\\
    \certCanAns(q, \tup{\A,\emptyset}, q') &= \sparqlAns(q, \A, q') \otimes \adm(q')\label{eq:6-0}
  \end{align}
  Then since $\sparqlAns(q, \A, q') \subseteq \sparqlAns(q', \A)$,
  for each $\omega \in \sparqlAns(q, \A, q')$,
  $\omega \in \sparqlAns(q', \A)$ must hold.\\
  So $\dom(\omega) \in \adm(q')$ must hold as well.\\
  Therefore $\{\omega|_X \mid X \in \max_\subseteq (\adm(q') \cap 2^{\dom(\omega)}\} = \{\dom(\omega)\}$.\\
  So from \ref{eq:6-0}:
  \begin{align}
    \certCanAns(q, \tup{\A,\emptyset}, q') &= \sparqlAns(q, \A, q')\label{eq:6-1}\\
\intertext{Finally:}
\certCanAns(q, \tup{\A,\emptyset}) &= \bigcup\limits_{q' \in \branch(q)} \certCanAns(q, \tup{\A,\emptyset}, q')\label{eq:6-2}\\
\intertext{So from~\ref{eq:6-1} and~\ref{eq:6-2}:}
\certCanAns(q, \tup{\A,\emptyset}) &= \bigcup\limits_{q' \in \branch(q)} \sparqlAns(q, \A, q')\\
\certCanAns(q, \tup{\A,\emptyset}) &= \bigcup\limits_{q' \in \branch(q)} (\sparqlAns(q, \A) \cap \sparqlAns(q', \A))\\
\certCanAns(q, \tup{\A,\emptyset}) &= \sparqlAns(q, \A) \cap \bigcup\limits_{q' \in \branch(q)} \sparqlAns(q', \A)\\
\intertext{And since:}
\sparqlAns(q, \tup{\A,\emptyset}) &\subseteq \bigcup\limits_{q'\ \in \branch(q)} \sparqlAns(q', \A)\\
\intertext{we get:}
\certCanAns(q, \tup{\A,\emptyset}) &= \sparqlAns(q, \A)
  \end{align}
\end{proof}

\begin{lemma}\label{lemma:certCanAns_optComp}
For any SUJO queries $q_1$, $q_2$ and  $\lcan$ KB $\K$:
$$\certCanAns(q_1, \K) \preceq_g \certCanAns(q_1\ \OPT\ q_2,\K)$$
\end{lemma}
 \begin{proof}
   Let $q_1,q_2$ be SUJO queries,
   let $\K$ be an $\lcan$ KB,
   and let $\omega_1 \in \certCanAns(q_1,\K)$.\\
    We need to show that there is an $\omega_2 \in \certCanAns(q_1\ \OPT\ q_2,\K)$ s.t. $\omega_1 \preceq \omega_2$.\\
    Since $\omega_1 \in \certCanAns(q_1,\K)$,
there must be an SJO query $q' \in \branch(q_1)$ s.t. $\omega_1 \in \certCanAns(q_1, \K, q')$.\\
So there is a $\rho_1 \in \sparqlAns(q_1, \can(\K)) \cap \sparqlAns(q', \can(\K))$ and an $X \in \max_\subseteq (\adm(q') \cap 2^{\dom(\rho_1)})$ s.t. $\omega_1 = \rho_1|_X$.\\
Since $\rho_1 \in \sparqlAns(q', \can(\K))$,
from Definition~\ref{def:sparqlAns},
there must be a $\rho_2 \in \sparqlAns(q'\ \OPT\ q_2, \can(\K))$ s.t. $\rho_1 \preceq \rho_2$.\\
We first show that 
$\rho_2 \in \sparqlAns(q_1\ \OPT\ q_2, \can(\K))$ must hold.\\
For this, we distinguish two cases:
\begin{itemize}
\item $\rho_1 = \rho_2$.\\
  From Definition~\ref{def:sparqlAns},
  for each $\rho_3 \in \sparqlAns(q_2,\can(\K))$,
  $\rho_1 \not\sim \rho_3$ must hold.\\
  Then because $\rho_1 \in \sparqlAns(q_1, \can(\K))$,
  from Definition~\ref{def:sparqlAns} still,
  $\rho_1 = \rho_2 \in \sparqlAns(q_1\ \OPT\ q_2, \can(\K))$ must hold. 
\item $\rho_1 \neq \rho_2$.\\
  Because
  $\rho_1 \in \sparqlAns(q', \can(\K))$,
  $\rho_2 \in \sparqlAns(q'\ \OPT\ q_2, \can(\K))$ and $\rho_1 \preceq \rho_2$,
  from Definition~\ref{def:sparqlAns},
  there must be a $\rho_3 \in \sparqlAns(q_2,\can(\K))$ s.t. $\rho_2 = \rho_1 \cup \rho_3$.\\
  So $\rho_1 \sim \rho_3$ holds.\\
  Then because $\rho_1 \in \sparqlAns(q_1, \can(\K))$,
  $\rho_3 \in \sparqlAns(q_2,\can(\K))$ and $\rho_1 \sim \rho_3$,
  from Definition~\ref{def:sparqlAns} still,
 $\rho_1 \cup \rho_3  = \rho_2 \in \sparqlAns(q_1\ \OPT\ q_2, \can(\K))$ must hold. 
\end{itemize}
\ \\
Now because $\rho_1 \preceq \rho_2$,
$\dom(\rho_1) = X \subseteq \dom(\rho_2)$.\\
And since $X \in \adm(q')$,
$X \in \adm(q') \cap 2^{\dom(\rho_2)}$ holds.\\
So there must be an $X'$ s.t. $X \subseteq X'$ and $X' \in \max_\subseteq (\adm(q') \cap 2^{\dom(\rho_2)})$.\\
Finally,
because $q' \in \branch(q_1)$,
from Definition~\ref{def:branch},
$q' \in \branch q_1\ \OPT\ q_2$.\\
So from Definition~\ref{def:certCanAns_SJO},
$ \rho_2|_{X'} \in \certCanAns(q_1\ \OPT\ q_2,\K)$.\\

\noindent Now let $\omega_2 = \rho_2|_{X'}$.\\
To complete the proof,
we only need to show that $\omega_1 \preceq \omega_2$.\\
First, since $\omega_1 = \rho_1|_X$,
$\omega_1 \preceq \rho_1$ must hold.\\
Then from the definition of $\rho_2$,
$\rho_1 \preceq \rho_2$.\\
So from the transitivity of  $\preceq$,
$\omega_1 \preceq \rho_2$.\\
Finally,
since $X \subseteq X'$,
$\omega_1|_X \preceq \rho_2|_{X'}$ must hold,
i.e. $\omega_1 \preceq \omega_2$.
\end{proof}
  
\begin{lemma}\label{lemma:certCanAns_vbComp}
  For any SUJO query $q$, $\lcan$ KB $\K$ and $\omega \in \certCanAns(q, \K)$:
  $$\dom(\omega) \in \adm(q)$$
\end{lemma}
\begin{proof}
Let  $q$ be a SUJO query and $\K$ an $\lcan$ KB.\\
Then $\certCanAns(q, \K) = \bigcup\limits_{q' \in \branch(q)} (\restRange{\sparqlAns(q, \A, q')}{\aDom(\K)}) \otimes \adm(q')$.\\
So for each $\omega \in \certCanAns(q, \K)$,
there is a $q' \in \branch(q)$ and solution mapping $\omega'$ s.t. $\omega = \omega'|_X$ for some $X \in \max_\subseteq (\adm(q') \cap 2^{\dom(\omega')})$.\\
So $\dom(\omega) \in \adm(q')$.\\
Then Lemma~\ref{lemma:branch_adm} below shows that for any $q' \in \branch(q)$,
$\adm(q') \subseteq \adm(q)$.\\
So $\dom(\omega) \in \adm(q')$.
\end{proof}

\begin{lemma}\label{lemma:branch_adm}
  For any SUJO query $q$ and PJO $q' \in \branch(q)$:
$$\adm(q') \subseteq \adm(q)$$
\end{lemma}
\begin{proof}
  Let $q$ be a SUJO query,
  $q' \in \branch(q)$ and $X \in \adm(q')$.\\
  We need to show that $X \in \adm(q)$.\\
  By induction on $q$:
    \begin{itemize}
      \item
        $q$ is a triple pattern.\\
        Then  $\branch(q) = \{q\}$, so the property trivially holds.
    \item  $q = \SELECT_Y\ q_2$.\\
      From Definition~\ref{def:branch},
      $q' = \SELECT_Y\ q_2'$ for some $q_2' \in \branch(q_2)$.\\
      So from Definition~\ref{def:adm_ind},
      $X = Y \cap Y'$ for some $Y' \in \adm(q_2')$.\\
      Then by IH,
      $Y' \in \adm(q_2)$.\\
      So $X = Y \cap Y'$ for some $Y' \in \adm(q_2)$.\\
      And again from Definition~\ref{def:adm_ind},
      $X \in \adm(\SELECT_Y\ q_2) = \adm(q)$.
    \item $q = q_1\ \JOIN\ q_2$.\\
      From Definition~\ref{def:branch},
      $q' = q'_1 \ \JOIN\ q'_2$ for some $(q'_1,q'_2) \in \branch(q_1) \times \branch(q_2)$.\\
      So from Definition~\ref{def:adm_ind},
      $X = X_1 \cup X_2$ for some  $(X_1,X_2) \in \adm(q'_1) \times \adm(q'_2)$.\\
      Then by IH,
      $X_1 \in \adm(q_1)$ and $X_2 \in \adm(q_2)$.\\
      So $X = X_1 \cup X_2$ for some  $(X_1,X_2) \in \adm(q_1) \times \adm(q_2)$.\\
      And again from Definition~\ref{def:adm_ind},
      $X \in \adm(q_1\ \JOIN\ q_2) = \adm(q)$.

    \item $q = q_1\ \UNION\ q_2$.\\
      From Definition~\ref{def:branch},
      $q' \in \branch(q_i)$ for some $i \in \{1,2\}$.\\
      So from Definition~\ref{def:adm_ind},
      $X \in \adm(q_i)$ for some $i \in \{1,2\}$.\\
      Then again from Definition~\ref{def:adm_ind},
      $X \in \adm(q_1\ \UNION\ q_2) = \adm(q)$.
    \item If $q = q_1\ \OPT\ q_2$,
      then $q' \in \branch(q_1\ \JOIN\ q_2)$ or $q' \in \branch(q_1)$ must hold.
      \begin{itemize}
      \item If $q' \in \branch(q_1\ \JOIN\ q_2)$,
        then we showed above that $X \in \adm(q_1\ \JOIN\ q_2)$ must hold.\\
      And from Definition~\ref{def:adm_ind},
      $\adm(q_1\ \JOIN\ q_2) \subseteq \adm(q)$.\\
      So $X \in \adm(q)$.
    \item  If $q' \in \branch(q_1)$,
      then by IH,
      $X \in \adm(q_1)$.\\
      And from Definition~\ref{def:adm_ind},
      $\adm(q_1) \subseteq \adm(q)$.\\
      So $X \in \adm(q)$.
      \end{itemize}
   \end{itemize}
\end{proof}

\begin{lemma}\label{lemma:certCanAns_provComp}
  For any queries $q_1, q_2$, $\lcan$ KB $\K$ and solution mapping $\omega$:
  $$\textnormal{if } \omega \in \certCanAns(q_1\ \UNION\ q_2) \textnormal{ and } \omega \not\in \certCanAns(q_2), \textnormal{ then }\dom(\omega) \in \adm(q_1)$$
\end{lemma}
\begin{proof}
Let $\omega \in \certCanAns(q_1\ \UNION\ q_2, \K)$ s.t. $\omega \not\in \certCanAns(q_2,\K)$.\\
Then from Definition~\ref{def:certCanAns},
because $\omega \in \certCanAns(q_1\ \UNION\ q_2, \K)$:
\begin{align*}
  \omega &\in  \bigcup\limits_{q' \in \branch(q_1\ \UNION\ q_2)} \certCanAns(q_1\ \UNION\ q_2, \K, q')\\
  \intertext{And from Definition~\ref{def:branch}:}
  \branch(q_1\ \UNION\ q_2) &= \branch(q_1) \cup \branch(q_2)\\
  \intertext{So:}
  \omega &\in  \bigcup\limits_{q' \in \branch(q_1) \cup \branch(q_2)} \certCanAns(q_1\ \UNION\ q_2, \K, q')\\
\end{align*}
So there is an SJO query $q' \in \branch(q_1) \cup \branch(q_2)$ s.t. $\omega \in \certCanAns(q_1\ \UNION\ q_2, \K, q')$\\
So there is an $\omega' \in \sparqlAns(q_1\ \UNION\ q_2, \can(\K)) \cap \sparqlAns(q', \can(\K))$ s.t.\\ $\omega = \omega'|_X$ for some $X \in \max_\subseteq (\adm(q') \cap 2^{\dom(\omega')})$.\\
Then we can distinguish three cases: 
\begin{itemize}
\item
  $q' \in \branch(q_1) \setminus \branch(q_2)$.\\
Since $\omega' \in \sparqlAns(q', \can(\K))$ and  $q' \not\in \branch(q_2)$,
$\omega' \not\in \sparqlAns(q_2, \can(\K))$ must hold.\\
Then because $\omega' \in \sparqlAns(q_1\ \UNION\ q_2, \can(\K))$,
from Definition~\ref{def:sparqlAns},
$\omega' \in \sparqlAns(q_1)$, must hold.\\
So  from Definition~\ref{def:certCanAns_SJO},
  $\omega \in \certCanAns(q_1,\K, q')$.\\
  And since $q' \in \branch(q_1)$,
  from Definition~\ref{def:certCanAns},
  $\omega \in \certCanAns(q_1,\K)$.\\
  So from Lemma~\ref{lemma:certCanAns_vbComp} above,
  $\dom(\omega) \in \adm(q_1)$
\item
  
$q' \in \branch(q_2) \setminus \branch(q_1)$.\\
Since $\omega' \in \sparqlAns(q', \can(\K))$ and  $q' \not\in \branch(q_1)$,
$\omega' \not\in \sparqlAns(q_1, \can(\K))$ must hold.\\
Then because $\omega' \in \sparqlAns(q_1\ \UNION\ q_2, \can(\K))$,
from Definition~\ref{def:sparqlAns},
$\omega' \in \sparqlAns(q_2)$, must hold.\\
So  from Definition~\ref{def:certCanAns_SJO},
  $\omega \in \certCanAns(q_2,\K, q')$.\\
  And since $q' \in \branch(q_2)$,
  from Definition~\ref{def:certCanAns},
  $\omega \in \certCanAns(q_2,\K)$,
  which would contradict the hypothesis.
\item
$q' \in \branch(q_1) \cap \branch(q_2)$.\\
Since $\omega' \in \sparqlAns(q_1\ \UNION\ q_2, \can(\K))$,
from Definition~\ref{def:sparqlAns},\\
$\omega' \in \sparqlAns(q_1, \can(\K))$ or $\omega' \in \sparqlAns(q_2, \can(\K))$ must hold.\\
If $\omega' \in \sparqlAns(q_1, \can(\K))$,
then from Definition~\ref{def:certCanAns_SJO},
  $\omega \in \certCanAns(q_1,\K, q')$.\\
  And since $q' \in \branch(q_1)$,
  from Definition~\ref{def:certCanAns},
  $\omega \in \certCanAns(q_1,\K)$.\\
  So from Lemma~\ref{lemma:certCanAns_vbComp} above,
  $\dom(\omega) \in \adm(q_1)$

  If $\omega' \in \sparqlAns(q_2, \can(\K))$ instead,
then from Definition~\ref{def:certCanAns_SJO},
  $\omega \in \certCanAns(q_2,\K, q')$.\\
  And since $q' \in \branch(q_2)$,
  from Definition~\ref{def:certCanAns},
  $\omega \in \certCanAns(q_2,\K)$,
  which would contradict the hypothesis.
\end{itemize}
\end{proof}

\section{Complexity proofs}
\label{sec:complexity_proofs}

\subsection{Proof of Proposition~\ref{prop:cost_adm}}
\label{sec:proof_base}
\propCostAdm*
\begin{proof}
  
Let $q$ be a JO query and $X_1, X_2 \subseteq \vars(q)$.\\

We reproduce here the inductive definition of $\base(q)$,
for readability.

\begin{definition}[Base of a JO query]\label{def:base}
   \begin{itemize}
   \item if $q$ is a triple pattern, then $\base(q) = \{\vars(q)\}$.
   \item if $q = q_1\ \JOIN\ q_2$, then $\base(q) = \{B_1 \cup B_2 \mid B_1 \in \min_\subseteq (\base(q_1)), B_2 \in \base(q_2)\} \cup$\\
     $\{B_1 \cup B_2 \mid B_1 \in \base(q_1), B_2 \in \min_\subseteq (\base(q_2))\}$
   \item if $q = q_1\ \OPT\ q_2$, then $\base(q) = \base(q_1) \cup \base(q_1\ \JOIN\ q_2)$
   \end{itemize}
   \end{definition}

In order to complete the proof sketched in Section~\ref{sec:complexity},
it is sufficient to show that:
\begin{itemize}
\item For any JO query $q$,
  the minimal element of $\base(q)$ w.r.t. set-inclusion is guaranteed to be unique.
This is shown with Lemma~\ref{lemma:base1} below.
\item $\adm(q) = \{\bigcup \B \mid \B \in 2^{\base(q)}\}$\label{item:6-2}.
This is shown with Lemma~\ref{lemma:base2} below.
\item $|\base(q)| = O(|q|)$\label{item:6-3}.
This is shown with Lemma~\ref{lemma:base3} below.
\end{itemize}
\end{proof}

\begin{lemma}\label{lemma:base1}
  For any JO query $q$, $|\min_\subseteq(\base(q))| = 1$.
\end{lemma}
\begin{proof}
  By induction on the structure of $q$.
  \begin{itemize}
  \item if $q$ is a triple pattern, then $|\base(q)| = 1$,
    so $|\min_\subseteq(\base(q))| = 1$.
  \item if $q = q_1\ \JOIN\ q_2$,
    let $\B_1 = \{B_1 \cup B_2 \mid B_1 \in \min_\subseteq (\base(q_1)), B_2 \in \base(q_2)\}$,
    and $\B_2 = \{B_1 \cup B_2 \mid B_1 \in \base(q_1), B_2 \in \min_\subseteq (\base(q_2))\}$.\\
    By IH,
    for $i \in \{1,2\}$,
    $|\min_\subseteq(\base(q_i))| = \{M_i\}$ for some $M_i \subseteq \vars(q_i)$.\\
    Then from the defintion of $\B_1$,
    $M_1 \cup M_2 \in \B_1$.\\
    And for each $B_2 \in \base(q_2)$,
    $M_2 \subseteq B_2$.\\
    So for each $M_1 \cup B_2 \in \B_1$,
    $M_1 \cup M_2 \subseteq M_1 \cup B_2$.\\
    So $\min_\subseteq (\B_1) = \{M_1 \cup M_2\}$.\\
    And similarly,
    $\min_\subseteq (\B_2) = \{M_1 \cup M_2\}$.\\
    Then because $\base(q) = \B_1 \cup \B_2$,
    $\min_\subseteq(\base(q)) = \{M_1 \cup M_2\}$.
  \item if $q = q_1\ \OPT\ q_2$,
    by IH,
    $\min_\subseteq(\base(q_1)) = \{M\}$ for some $M \subseteq \vars(q_1)$.\\
    So $M \subseteq B$ for each $B \in \base(q_1)$.\\
    And we showed above that $M \subseteq B$ for each $B \in \base(q_1\ \JOIN\ q_2)$.\\
    Then from Definitio~\ref{lemma:base1},
    $\base(q) = \base(q_1) \cup \base(q_1\ \JOIN\ q_2)$.\\
    So $M \in \base(q_1) \subseteq \base(q)$,
    and $M \subseteq B$ for each $B \in \base(q_1) \cup \base(q_1\ \JOIN\ q_2) = \base(q)$.\\
    Therefore $\min_\subseteq(\base(q)) = \{M\}$.
  \end{itemize}
\end{proof}
\begin{lemma}\label{lemma:base2}
  For any JO query $q$,
  $\adm(q) = \{\bigcup \B \mid \B \in 2^{\base(q)}\}$
  
\end{lemma}
\begin{proof}
  By induction on the structure of $q$.
  \begin{itemize}
  \item if $q$ is a triple pattern, then $\base(q) = \adm(q) = \{\vars(q)\}$.
    
  \item if $q = q_1\ \JOIN\ q_2$:
    \begin{itemize}
    \item ($\Rightarrow$).\\
      Let $X \in \adm(q)$.\\
      From Definition~\ref{def:adm_ind},
      $X = X_1 \cup X_2$ for some $(X_1,X_2) \in \adm(q_1) \times \adm(q_2)$.\\
      And by IH,
      for $i \in \{1,2\}$,
      $X_i = \bigcup \B_i$ for some $\B_i \in 2^{\base(q_1)}$.\\
      Then from Lemma~\ref{lemma:base1},
      $|\min_\subseteq(\base(q_i))| = \{M_i\}$ for some $M_i \subseteq \vars(q_i)$.\\
      So for each $B_i \in \B_i$,
      $M_i \subseteq B_i$.\\
      Therefore $\bigcup \B_i = \{M_i\} \cup \bigcup \B_i$.\\
      And since $X = X_1 \cup X_2$, we have:
      \begin{align*}
      X =& \bigcup \B_1 \cup \bigcup \B_2\\
      X =& \{M_1\} \cup \bigcup \B_1 \cup \{M_2\} \cup \bigcup \B_2\\
      X =& \{M_2 \cup B_1 \mid B_1 \in \B_1\} \cup \{M_1 \cup B_2 \mid B_2 \in \B_2\}
      \end{align*}
      Then from Definition~\ref{def:base},
      for each $B_1 \in \B_1$,
      $M_2 \cup B_1 \in \base(q)$.\\
      Similarly, for each $B_2 \in \B_2$,
      $M_1 \cup B_2 \in \base(q)$.\\
      So $X = \bigcup \B$ for some $\B \in 2^{\base(q)}$.
    \item ($\Leftarrow$).\\
      Let $X = \bigcup \B$  for some $\B \in 2^{\base(q)}$.\\
      From Definition~\ref{def:base},
      for each $B \in \B$,
      there are $(B_1,B_2) \in \base(q_1) \times \base(q_2)$ s.t. $B = B_1 \cup B_2$.\\
      For $i \in \{1,2\}$, let $\B_i = \{B_i \mid B_i \cup B' \in \B, B_i \in \base(q_i)\}$.\\
      Then for $i \in \{1,2\}$, $\B_i \neq \emptyset$.\\
      And $\B = \B_1 \cup \B_2$.\\
    So $X = \bigcup \B = \bigcup \B_1 \cup \bigcup \B_2$.\\
    And by IH,
    for $i \in \{1,2\}$,
  $\bigcup \B_i \in \adm(q_i)$.\\
  Therefore  $X = X_1 \cup X_2$ for some $(X_1,X_2) \in \adm(q_1) \times \adm(q_2)$.\\
  So From Definition~\ref{def:adm_ind},
      $X \in \adm(q)$.
    \end{itemize}
  \item if $q = q_1\ \OPT\ q_2$:
    \begin{itemize}
    \item ($\Rightarrow$).\\
      Let $X \in \adm(q)$.\\
      From Definition~\ref{def:adm_ind},
      $X \in \adm(q_1)$ or
      $X \in \adm(q_1\ \JOIN\ q_2 )$ must hold.\\
      If $X \in \adm(q_1)$,
      then by IH,
      $X = \bigcup \B$  for some $\B \in 2^{\base(q_1)}$.\\
      And from Definition~\ref{def:base},
      $\base(q_1) \subseteq \base(q)$.\\
      If $X \in \adm(q_1\ \JOIN\ q_2 )$,
      then we showed above that $X = \bigcup \B$  for some $\B \in 2^{\base(q_1\ \JOIN\ q_2)}$.\\
       And from Definition~\ref{def:base},
       $\base(q_1\ \JOIN\ q_2) \subseteq \base(q)$.\\
       So in both cases, $X = \bigcup \B$  for some $\B \in 2^{\base(q)}$.
      
     \item ($\Leftarrow$).\\
       Let $X = \bigcup \B$  for some $\B \in 2^{\base(q)}$.\\
       From Definition~\ref{def:base},
       for each $B \in \B$,
       $B \in \base(q_1)$ or there are $(B_1,B_2) \in \base(q_1) \times \base(q_2)$ s.t. $B = B_1 \cup B_2$.\\
       For $i \in \{1,2\}$, let $\B_i = \{B_i \mid B_i \cup B' \in \B, B_i \in \base(q_i)\}$.\\
       Then $\B_1 \neq \emptyset$.\\
       And $\B = \B_1 \cup \B_2$.\\

       \noindent If $\B_2 = \emptyset$,
       then $X = \bigcup \B = \bigcup \B_1$.\\
       And by IH,
       $\bigcup \B_1 \in \adm(q_1)$.\\
       So From Definition~\ref{def:adm_ind},
       $X \in \adm(q)$.\\

       \noindent If $\B_2 \neq \emptyset$,
       then $X = \bigcup \B = \bigcup \B_1 \cup \bigcup \B_2$.\\
       And by IH,
       for $i \in \{1,2\}$,
       $\bigcup \B_i \in \adm(q_i)$.\\
       Therefore  $X = X_1 \cup X_2$ for some $(X_1,X_2) \in \adm(q_1) \times \adm(q_2)$.\\
       So From Definition~\ref{def:adm_ind},
      $X \in \adm(q)$.
  \end{itemize}
  \end{itemize}
\end{proof}

\begin{lemma}\label{lemma:base3}
  For any JO query $q$, $|\base(q)| = O(|q|)$
\end{lemma}
\begin{proof}
  By induction on the structure of $q$.
  \begin{itemize}
  \item if $q$ is a triple pattern, then $|\base(q)| = 1$.
  \item if $q = q_1\ \JOIN\ q_2$,
     then immediately from the definition of $\base(q)$,\\
     $|\base(q)| = O(
     |\min_\subseteq(\base(q_1))|)
     \cdot 
     |\base(q_2)|
     +
     |\min_\subseteq(\base(q_2))|)
     \cdot 
     |\base(q_1)|)$.\\
     So from Lemma~\ref{lemma:base1},
    $|\base(q)| = O(|\base(q_2)| + |\base(q_1)|)$.\\
    And by IH,
    $|\base(q_i)| = O(|q_i|)$ for $i \in \{1,2\}$.\\
  So $|\base(q)| = O(|q_1|) + O(|q_2|) = O(|q|)$.
 \item if $q = q_1\ \OPT\ q_2$,
   the argument is similar to the case $q = q_1\ \JOIN\ q_2$.
  \end{itemize}
\end{proof}

\end{document}